\PassOptionsToPackage{unicode}{hyperref}
\PassOptionsToPackage{hyphens}{url}
\PassOptionsToPackage{dvipsnames,svgnames,x11names}{xcolor}
\documentclass[
]{article}

\usepackage{amsmath,amssymb}
\usepackage{iftex}
\ifPDFTeX
  \usepackage[T1]{fontenc}
  \usepackage[utf8]{inputenc}
  \usepackage{textcomp} 
\else 
  \usepackage{unicode-math}
  \defaultfontfeatures{Scale=MatchLowercase}
  \defaultfontfeatures[\rmfamily]{Ligatures=TeX,Scale=1}
\fi
\usepackage{lmodern}
\ifPDFTeX\else  
\fi
\IfFileExists{upquote.sty}{\usepackage{upquote}}{}
\IfFileExists{microtype.sty}{
  \usepackage[]{microtype}
  \UseMicrotypeSet[protrusion]{basicmath} 
}{}
\makeatletter
\@ifundefined{KOMAClassName}{
  \IfFileExists{parskip.sty}{%
    \usepackage{parskip}
  }{
    \setlength{\parindent}{0pt}
    \setlength{\parskip}{6pt plus 2pt minus 1pt}}
}{
  \KOMAoptions{parskip=half}}
\makeatother
\usepackage{xcolor}
\ifx\paragraph\undefined\else
  \let\oldparagraph\paragraph
  \renewcommand{\paragraph}[1]{\oldparagraph{#1}\mbox{}}
\fi
\ifx\subparagraph\undefined\else
  \let\oldsubparagraph\subparagraph
  \renewcommand{\subparagraph}[1]{\oldsubparagraph{#1}\mbox{}}
\fi

\usepackage{longtable,booktabs,array}
\usepackage{calc} 
\usepackage{etoolbox}
\makeatletter
\patchcmd\longtable{\par}{\if@noskipsec\mbox{}\fi\par}{}{}
\makeatother
\IfFileExists{footnotehyper.sty}{\usepackage{footnotehyper}}{\usepackage{footnote}}
\makesavenoteenv{longtable}
\usepackage{graphicx}
\makeatletter
\def\maxwidth{\ifdim\Gin@nat@width>\linewidth\linewidth\else\Gin@nat@width\fi}
\def\maxheight{\ifdim\Gin@nat@height>\textheight\textheight\else\Gin@nat@height\fi}
\makeatother
\setkeys{Gin}{width=\maxwidth,height=\maxheight,keepaspectratio}
\makeatletter
\def\fps@figure{htbp}
\makeatother

\usepackage{arxiv}
\usepackage{orcidlink}
\usepackage{amsmath}
\usepackage[T1]{fontenc}
\makeatletter
\@ifpackageloaded{caption}{}{\usepackage{caption}}
\AtBeginDocument{%
\ifdefined\contentsname
  \renewcommand*\contentsname{Table of contents}
\else
  \newcommand\contentsname{Table of contents}
\fi
\ifdefined\listfigurename
  \renewcommand*\listfigurename{List of Figures}
\else
  \newcommand\listfigurename{List of Figures}
\fi
\ifdefined\listtablename
  \renewcommand*\listtablename{List of Tables}
\else
  \newcommand\listtablename{List of Tables}
\fi
\ifdefined\figurename
  \renewcommand*\figurename{Figure}
\else
  \newcommand\figurename{Figure}
\fi
\ifdefined\tablename
  \renewcommand*\tablename{Table}
\else
  \newcommand\tablename{Table}
\fi
}
\@ifpackageloaded{float}{}{\usepackage{float}}
\floatstyle{ruled}
\@ifundefined{c@chapter}{\newfloat{codelisting}{h}{lop}}{\newfloat{codelisting}{h}{lop}[chapter]}
\floatname{codelisting}{Listing}

\makeatother
\makeatletter
\@ifpackageloaded{caption}{}{\usepackage{caption}}
\@ifpackageloaded{subcaption}{}{\usepackage{subcaption}}
\makeatother
\makeatletter
\@ifpackageloaded{tcolorbox}{}{\usepackage[skins,breakable]{tcolorbox}}
\makeatother
\makeatletter
\@ifundefined{shadecolor}{\definecolor{shadecolor}{rgb}{.97, .97, .97}}
\makeatother
\ifLuaTeX
  \usepackage{selnolig}  
\fi
\usepackage[round]{natbib}
\IfFileExists{bookmark.sty}{\usepackage{bookmark}}{\usepackage{hyperref}}
\IfFileExists{xurl.sty}{\usepackage{xurl}}{} 
\hypersetup{
  pdftitle={Combining support for hypotheses over heterogeneous studies with Bayesian Evidence Synthesis: A simulation study},
  pdfauthor={Thom Benjamin Volker; Irene Klugkist},
  colorlinks=true,
  linkcolor={blue},
  filecolor={Maroon},
  citecolor={Blue},
  urlcolor={Blue},
  pdfcreator={LaTeX via pandoc}}

\renewcommand{\today}{2023-12-22}
\newcommand{\runninghead}{A Preprint }
\renewcommand{\runninghead}{Bayesian Evidence Synthesis }
\title{Combining support for hypotheses over heterogeneous studies with
Bayesian Evidence Synthesis: A simulation study}
\def\asep{\\\\\\ } 
\def\asep{\And }
\author{\textbf{Thom Benjamin
Volker}~\orcidlink{0000-0002-2408-7820}\\Department of Methodology and
Statistics\\Utrecht University\\Utrecht,\ 3584
CH\\\href{mailto:t.b.volker@uu.nl}{t.b.volker@uu.nl}\asep\textbf{Irene
Klugkist}~\orcidlink{0000-0001-9561-3691}\\Department of Methodology and
Statistics\\Utrecht University\\Utrecht,\ 3584
CH\\\href{mailto:i.klugkist@uu.nl}{i.klugkist@uu.nl}}
\date{2023-12-22}
\begin{document}
\maketitle
\begin{abstract}
Scientific claims gain credibility by replicability, especially if
replication under different circumstances and varying designs yields
equivalent results. Aggregating results over multiple studies is,
however, not straightforward, and when the heterogeneity between studies
increases, conventional methods such as (Bayesian) meta-analysis and
Bayesian sequential updating become infeasible. \emph{Bayesian Evidence
Synthesis}, built upon the foundations of the Bayes factor, allows to
aggregate support for conceptually similar hypotheses over studies,
regardless of methodological differences. We assess the performance of
Bayesian Evidence Synthesis over multiple effect and sample sizes, with
a broad set of (inequality-constrained) hypotheses using Monte Carlo
simulations, focusing explicitly on the complexity of the hypotheses
under consideration. The simulations show that this method can evaluate
complex (informative) hypotheses regardless of methodological
differences between studies, and performs adequately if the set of
studies considered has sufficient statistical power. Additionally, we
pinpoint challenging conditions that can lead to unsatisfactory results,
and provide suggestions on handling these situations. Ultimately, we
show that Bayesian Evidence Synthesis is a promising tool that can be
used when traditional research synthesis methods are not applicable due
to insurmountable between-study heterogeneity.
\end{abstract}
\ifdefined\Shaded\renewenvironment{Shaded}{\begin{tcolorbox}[frame hidden, sharp corners, boxrule=0pt, interior hidden, borderline west={3pt}{0pt}{shadecolor}, breakable, enhanced]}{\end{tcolorbox}}\fi

\hypertarget{introduction}{%
\section{Introduction}\label{introduction}}

In recent years, a meta-analytic way of thinking has been advocated in
the scientific community. This approach is grounded in the belief that a
single study is merely contributing to a larger body of evidence
\citep[e.g.,][]{asendorpf_recommendations_2016, cumming_new_2014, goodman_reproducibility_2016}.
Such evidence gains credibility only by replicability of the findings
with new data \citep{schmidt_replication_2009}, because the ability to
replicate research findings ensures that the findings represent
\emph{true} phenomena rather than artefacts. The replication crisis in
the social sciences placed the importance of replication back on the
research agenda and kickstarted multiple replication initiatives.
Accordingly, multiple efforts aimed at fostering replication were
undertaken, such as journals or journal sections devoted to replication
studies (e.g., Royal Society Open Science, Registered Replication
Reports, Journal of Personality and Social Psychology) and grant
opportunities for replication studies
\citep[e.g.,][]{nwo_replication_2020}. Moreover, multiple scholars
legitimately emphasized the importance of replication studies for the
future of science
\citep[e.g.,][]{baker_reproducibility_2016, brandt_et_al_replication_2014, munafo_manifesto_2017}.

This renewed interest in replication was mostly directed toward studies
that are highly similar, using a methodology and research design that
mimics the original study as closely as possible
\citep[e.g.,][]{camerer2016evaluating, camerer2018evaluating, klein_etal_replicability_2014, nosek_replicability_review_2021, open_science_collab_2015}.
These studies, commonly referred to as \emph{direct}, \emph{exact} or
\emph{close} replications, are primarily concerned with the statistical
reliability of the results. As such, direct replications are tailored
towards assessing whether or not the results of the initial study are
due to chance. Agreement between the findings of a direct replication
and the findings of the initial study increases the confidence in the
accuracy of the original findings. However, direct replicability is a
necessary but insufficient condition for making scientific claims. If
the results of the studies depend on methodological flaws, inferences
from all studies will lead to suboptimal or invalid conclusions
\citep{lawlor_triangulation_2017, munafo_robust_2018}.

\emph{Conceptual} replications protect against placing too much
confidence in findings that depend on methodological shortcomings. A
conceptual replication primarily assesses the validity and
generalizability of a study, by testing whether similar results can be
obtained under different circumstances, or using different methods and
operationalizations \citep{nosek_scientific_2012}. The rationale is that
a phenomenon that can be observed in a variety of research settings is
more likely to represent a \emph{true} effect than a finding that
replicates only under particular circumstances
\citep{crandall_conceptual_2016}. Additionally, different methodologies
used in different studies may have different strengths and weaknesses,
that may all affect the conclusions drawn from the data. Combining
evidence from multiple approaches mitigates the effect of these
strengths and weaknesses, and thereby enhances the validity and the
robustness of the final conclusion
\citep{lawlor_triangulation_2017, lipton2003inference, mathison1988triangulate, munafo_robust_2018, nosek_scientific_2012}.

In the conventional framework of direct replications, combining evidence
over studies is relatively straightforward, because established methods
as (Bayesian) meta-analysis
\citep{lipsey_wilson_2001, sutton_bayesian_meta2001} or Bayesian
sequential updating \citep{schonbrodt_sequential_2017} can be applied to
aggregate the results. These methods pool the parameter estimates or
effect sizes obtained in the individual studies
\citep{cooper_handbook_2009}. Even if the studies are not identical, but
still considerably similar in terms of study design and analysis
methods, meta-analysis can be used, and moderators can be added to
explain variability in the estimated effects. However, if the studies
differ considerably with regard to research design, operationalizations
of key variables or statistical models used, the parameter estimates or
effect sizes will not be comparable. For example, the set of studies may
include both experimental and correlational studies, varying covariates,
or use different operationalizations of the same construct (e.g.,
different measurement instruments or different tasks). Under such
circumstances, aggregating the varying effect sizes may not be
meaningful, which renders the use of these conventional approaches
infeasible.

To overcome this problem, \citet{kuiper_combining_2013} proposed a new
method called \emph{Bayesian Evidence Synthesis} (\emph{BES}). The key
of the approach is to aggregate the \emph{evidence} for a scientific
theory or overarching hypothesis, rather than effect sizes, by
aggregating Bayes factors obtained in individual studies. In every
single study, a statistical hypothesis can be formulated that reflects
the overall theory, but that accounts for characteristics of the data
and research methodology unique to that study. The evidence for each
study-specific hypothesis can be expressed using a Bayes factor
(\(BF\)), rendering the relative support for the hypothesis of interest
over some alternative hypothesis
\citep{kass_raftery_bayes_factors_1995}. If the study-specific
hypotheses reflect the same underlying theory, the effect sizes might be
too heterogeneous to aggregate, but the support for the theory,
quantified using Bayes factors, can be meaningfully combined. After all,
each individual study provides a certain amount of evidence for or
against this overarching theory. The evidence in each study can be
aggregated into a single measure that reflects the support for the
theory over all studies combined.

The popularity of Bayesian statistics in general
\citep[e.g.,][]{lynch_bayesian_2019} and Bayesian hypothesis evaluation
specifically \citep{vandeschoot_systematic_2017} rapidly increased over
recent years, especially in the context of individual studies. In
contrast to the classical null hypothesis testing paradigm, Bayesian
hypothesis evaluation enables researchers to compare potentially
non-nested hypotheses with (multiple) equality- and/or
order-constraints, by balancing the fit of the hypotheses to the data
with the complexity of the candidate hypotheses
\citep{klugkist_inequality_2005, hoijtink2019tutorial}. Moreover, the
Bayesian framework allows to quantify the amount of evidence from the
data for or against the hypothesis of interest, rather than merely
assessing whether the classical null hypothesis can be rejected
\citep{Wagenmakers_bayesian_2018}. Yet, although the theoretical
foundations of \emph{BES} have been laid out
\citep{kuiper_combining_2013}, and the method has been applied
successfully in substantive research
\citep[e.g.,][]{kevenaar_bes_2021, zondervan_parental_2019, zondervan_robust_2020, volker_cooperation_2022},
methodological research on Bayesian evaluation of hypotheses has hardly
been extended to the case in which evidence is combined over multiple
studies. It is not obvious how research on Bayesian hypothesis
evaluation within studies translates to the situation in which evidence
is combined over multiple studies. As a consequence, the required
conditions for adequate performance of \emph{BES} are unknown.

In this paper, we assess the performance of \emph{BES} in various
scenarios relevant to applied researchers. In multiple simulations, we
apply \emph{BES} to aggregate the evidence for conceptually similar
hypotheses that are evaluated using different statistical models, while
varying the operationalizations of key variables in these hypotheses. We
restrict the simulations to the evaluation of regression coefficients in
three members of the generalized linear model family: ordinary least
squares (OLS), logistic and probit regression, because of their
widespread appearance in empirical research. Note, however, that the
applicability of \emph{BES} reaches beyond these methods: as long as
conceptually similar hypotheses can be evaluated using Bayes factors,
\emph{BES} can be used to aggregate the results.

This flexibility may pose challenges to the application of \emph{BES}.
Different operationalizations of key variables may affect the
complexities of the hypotheses evaluated in the individual studies. As
Bayes factors directly depend on the complexities of the hypotheses, the
risk is that the resulting amount of evidence largely depends on the
specification of the hypotheses, rather than on their veracity. Our
simulations assess how common procedures in data handling affect the
complexities of the hypotheses and the resulting performance of
\emph{BES}. Additionally, we investigate in which situations \emph{BES}
does not perform adequately, and provide suggestions on how such
situations can be handled. Yet, before discussing the simulations, a
detailed description of \emph{BES} tailored to the specification of the
simulation study is provided. The paper concludes with an extensive
discussion of the implications of the simulations.

\hypertarget{aggregating-evidence-with-bes}{%
\section{\texorpdfstring{Aggregating Evidence with
\emph{BES}}{Aggregating Evidence with BES}}\label{aggregating-evidence-with-bes}}

Using \emph{BES} to quantify the evidence for a scientific theory
requires three building blocks. First, \emph{BES} requires multiple
studies on the same phenomenon, that all assess a conceptually similar
hypothesis \citep{kuiper_combining_2013}. Hence, in each study, a
hypothesis must be formulated that reflects this scientific theory, but
that also accounts for the specifics of the study, such as the empirical
design, the analysis model and operationalizations of key variables.
Second, these hypotheses are evaluated in each study separately. In the
Bayesian framework, this can be done by calculating the Bayes factor for
an hypothesis over an alternative hypothesis, using the posterior
distribution of the model parameters. Third, the support for the theory
in each individual study is aggregated using an updating procedure that
renders the support for the theory over all studies combined. Each step
is outlined in more detail in this section.

\hypertarget{informative-hypotheses}{%
\subsection{Informative hypotheses}\label{informative-hypotheses}}

When translating a conceptual hypothesis (i.e., the scientific theory)
to a statistical hypothesis, researchers conventionally relied on the
null hypothesis testing framework. In the context of regression models,
this generally implies that researchers evaluate the null hypothesis,
\(H_0\), stating that the regression coefficients \(\beta_k\) (where
\(k\) is an index) equal zero, against an unconstrained alternative
hypothesis, \(H_u\), indicating that they can have any value. Several
authors emphasized shortcomings of this approach, remarking that the
ability to reject a null hypothesis is hardly illuminating, because the
null is seldom (exactly) true
\citep{cohen_earth_1994, lykken_wrong_1991} and rarely corresponds to
the expectations researchers have
\citep{cohen_things_i_learned_1990, vandeschoot_informative_2011, royall1997statistical, trafimow_manipulating_2018}.
Multiple researchers argued that scientific expectations can be better
captured in an informative hypothesis
\citep{hoijtink_informative_2012, vandeschoot_informative_2011}.

Informative hypotheses allow to formalize the expectations researchers
have on the parameters of the statistical model
\citep{hoijtink_informative_2012}. Rather than expecting that the
regression coefficients under evaluation are equal to zero, researchers
may, for instance, expect that all are positive. In a regression model
with three predictors, this expectation yields \[
H_i: \{\beta_1, \beta_2, \beta_3\} > 0.
\] However, informative hypotheses are not restricted to
inequality-constrained hypotheses. An informative hypothesis might
contain equality and inequality constraints between parameters and
between combinations of parameters, as well as expectations regarding
effect sizes or range constraints \citep{hoijtink2019tutorial}.
Combining these features yields great flexibility to researchers to
formalize their theories, potentially resulting in fairly complex
hypotheses. For example, the expectation that \(\beta_1\) and
\(\beta_2\) are of the same size, while both are positive but smaller
than \(\beta_3\), can be formalized as \[
H_{i'}: 0 < \{\beta_1=\beta_2\} < \beta_3. 
\] The flexibility with regard to specifying hypotheses is especially
advantageous in the context of \emph{BES}, where studies may differ in
the operationalizations of key variables, and thus in the hypotheses
that are evaluated.

\hypertarget{hypothesis-evaluation-using-the-bayes-factor}{%
\subsection{Hypothesis evaluation using the Bayes
factor}\label{hypothesis-evaluation-using-the-bayes-factor}}

Within the Bayesian framework, the support for (informative) hypotheses
can be expressed in terms of a Bayes factor
\citep{kass_raftery_bayes_factors_1995}. Bayes factors require that a
likelihood function (i.e., an analysis model for the data) is combined
with a prior distribution, resulting in the posterior distribution of
the parameters. The prior and posterior distribution can subsequently be
used to calculate the Bayes factor.

\hypertarget{likelihood-of-the-parameters}{%
\subsubsection{Likelihood of the
parameters}\label{likelihood-of-the-parameters}}

After formulating the hypotheses of interest, researchers need to
specify a statistical model (i.e., a likelihood function) that allows to
evaluate these hypotheses when analyzing the data
\citep{lynch_introduction_2007}. A likelihood function quantifies the
support in the data for each potential parameter value. This paper is
restricted to OLS, logistic and probit regression, which are all part of
the generalized linear model (GLM) family (note that there are many
other models that fall under the GLM family). GLMs extend the linear
regression model to deal with response variables that are assumed to
follow a distribution from the exponential family. This is done by using
a link function
\(g(\boldsymbol{\mu}) = \boldsymbol{X} \boldsymbol{\beta}\), where
\(\boldsymbol{X}\) is an \(n \times p\) matrix containing \(n\)
observations' scores on \(p\) predictor variables including an indicator
for the intercept, \(\boldsymbol{\beta}\) is a \(p \times 1\) vector
containing the regression coefficients including intercept, and
\(\boldsymbol{\mu} = E(\boldsymbol{y}|\boldsymbol{X})\), where
\(\boldsymbol{y}\) is an \(n \times 1\) vector with response values. The
purpose of the link function is to model each observation's expected
response \(E(\boldsymbol{y}|\boldsymbol{X})\) in terms of a linear
combination of the predictor variables. Accordingly, the likelihood of
the parameters given the data under the generalized linear model is
formalized as \[
L(\boldsymbol{\beta}, \phi| \boldsymbol{y}, \boldsymbol{X}) \equiv 
f(\boldsymbol{y}|\boldsymbol{X}, \boldsymbol{\beta}, \phi) = 
\prod^n_{i=1} f(y_i|\boldsymbol{x}_i, \boldsymbol{\beta}, \phi),
\] where \(i\) indexes each observation and \(\phi\) denotes the
variance or dispersion parameter.

One of the best known models in the class of GLMs is the linear
regression model, which uses the identity link function, such that
\(g(\boldsymbol{\mu}) = \boldsymbol{\mu} = \boldsymbol{X\beta}\). The
likelihood function of the linear regression model is defined by \[
f(\boldsymbol{y} | \boldsymbol{X}, \boldsymbol{\beta}, \phi) = \prod^n_{i=1} \frac{1}{\sqrt{2\pi\sigma^2}} \text{exp}
\Bigg\{
- \frac{(y_i - \boldsymbol{x}_i\boldsymbol{\beta})^2}{2\sigma^2}
\Bigg\},
\] where \(\phi = \sigma^2\). When the outcome \(\boldsymbol{y}\) is
dichotomous rather than continuous, the binomial distribution can be
used. The corresponding likelihood is defined by \[
f(\boldsymbol{y}|\boldsymbol{X}, \boldsymbol{\beta}) = \prod^n_{i=1} p_i^{y_i} (1 - p_i)^{1 - y_i},
\] where \(y_i\) is either \(0\) or \(1\), and \(p_i\) is specified
according to the logit link
\(g(\boldsymbol{\mu}) = \log\Big(\frac{\boldsymbol{\mu}}{1 - \boldsymbol{\mu}}\Big) = \boldsymbol{X\beta}\)
for logistic regression, or probit link
\(g(\boldsymbol{\mu}) = \Phi^{-1}(\boldsymbol{\mu}) = \boldsymbol{X\beta}\),
where \(\Phi^{-1}\) denotes the inverse cumulative distribution function
of the standard normal distribution, for probit regression models. The
dispersion parameter is omitted because it equals \(\phi = 1\) for
logistic and probit models.

\hypertarget{prior-distribution}{%
\subsubsection{Prior distribution}\label{prior-distribution}}

After specifying the analysis model, the prior distribution for the
model parameters must be defined. Because the prior for an informative
hypothesis can be obtained by truncating the unconstrained (or
encompassing) prior \citep[e.g.,][]{klugkist_inequality_2005}, we first
discuss the latter. The prior distribution reflects which parameter
values are most likely before observing the data, and hence may be more
or less informative depending on the amount of prior knowledge
researchers possess. Although researchers have substantial freedom in
specifying the prior distribution, inadequate choices have adverse
consequences for hypothesis evaluation and comparison, rendering it a
complicated task \citep{ohagan_fractional_1995}. A practical solution to
the specification of the prior for hypothesis evaluation, is to use a
fractional unconstrained prior \citep{ohagan_fractional_1995} that
protects against a subjective specification
\citep{gu_approximated_2018}. This approach entails that a
non-informative improper prior \(p_u(\boldsymbol{\beta}, \phi)\) is
updated with a small fraction \(b\) of the information in the data
(i.e., the likelihood), resulting in a proper default prior distribution
\citep[e.g.,][]{gu_approximated_2018, mulder_equality_2010}, such that
\[
p_u(\boldsymbol{\beta}, \phi|\boldsymbol{y}, \boldsymbol{X}, b) = 
\frac{
  p_u(\boldsymbol{\beta}, \phi) ~ f(\boldsymbol{y}|\boldsymbol{X}, \boldsymbol{\beta}, \phi)_b
}{
  \int \int p_u(\boldsymbol{\beta}, \phi) ~ 
  f(\boldsymbol{y}|\boldsymbol{X}, \boldsymbol{\beta}, \phi)_b ~ 
  \partial \boldsymbol{\beta} \partial \phi
}.
\] \citet{mulder_olssoncollentine_2019} proved that, under the linear
model, updating a noninformative improper Jeffrey's prior with a
fraction \(b = \frac{p+1}{n}\) of the likelihood renders a Cauchy
distribution (i.e., a Student \(\mathcal{T}\) prior distribution with 1
degree of freedom) for the regression coefficients, \[
p_u(\boldsymbol{\beta} | \boldsymbol{y}, \boldsymbol{X}, b) = 
\mathcal{T}(\boldsymbol{\hat{\beta}}, \boldsymbol{\hat{\Sigma}_\beta} / b, 1),
\] with location parameter \(\boldsymbol{\hat{\beta}}\) (the maximum
likelihood estimates of the regression coefficients) and scale parameter
\(\boldsymbol{\hat{\Sigma}_{\beta}} / b\) (the unbiased estimate of the
covariance matrix of the regression coefficients over \(b\)). A
discussion of the prior for the dispersion parameter is omitted, because
nuisance parameters are integrated out when calculating the Bayes factor
\citep{gu_approximated_2018}.

When the model under evaluation is not (multivariate) normal, but
another GLM, obtaining an analytically tractable posterior may be
difficult, regardless of the choice of prior \citep{bda2013}.
Nevertheless, large-sample theory dictates that a fractional prior and
posterior for the regression coefficients can be approximated by a
multivariate normal distribution \citep{bda2013}. In this case,
\citet{gu_approximated_2018} showed how a fractional prior can be
obtained by updating a non-informative prior with a fraction
\(b = \frac{J}{n}\) of the likelihood. This approach renders an
approximately normal prior distribution for the regression coefficients
\[
p_u(\boldsymbol{\beta} | \boldsymbol{y}, \boldsymbol{X}, b) \approx 
\mathcal{N}(\boldsymbol{\hat{\beta}}, \boldsymbol{\hat{\Sigma}_\beta} / b).
\] with mean vector \(\boldsymbol{\hat{\beta}}\) and covariance matrix
\(\boldsymbol{\hat{\Sigma}_{\beta}}/b\). Typically, \(J\) equals the
number of independent constraints in all hypotheses of interest within a
study, but different specifications are possible \citep[for an elaborate
discussion on appropriate values for \(J\),
see][]{gu_approximated_2018, hoijtink_prior_2021}.

After specifying a prior for the unconstrained hypothesis, the
encompassing prior approach can be used to obtain a prior for an
informative hypothesis by truncating the unconstrained prior
\citep[e.g.,][]{klugkist_inequality_2005, mulder_equality_2010, mulder_prior_2014}.
The prior for an informative hypothesis is proportional to the region of
the unconstrained prior that is in agreement with the constraints
imposed by the hypothesis under consideration. That is,
\(p_i(\boldsymbol{\beta} | \boldsymbol{y}, \boldsymbol{X}, b) \propto p_u(\boldsymbol{\beta} | \boldsymbol{y}, \boldsymbol{X}, b)\boldsymbol{1}_{\boldsymbol{B}_i}\),
where \(\boldsymbol{1}_{\boldsymbol{B}_i}\) is an indicator for the
admissible parameter space \(\boldsymbol{B}_i\) under hypothesis \(H_i\)
\citep{gu_approximated_2018}.

\hypertarget{posterior-distribution}{%
\subsubsection{Posterior distribution}\label{posterior-distribution}}

Combining the likelihood of the parameters with the prior distribution
yields the posterior distribution. The posterior quantifies the support
for each parameter value after observing the data. Under the fractional
prior approach, the unconstrained fractional prior is updated with the
remaining fraction \(1-b\) of the likelihood. This approach yields the
posterior distribution under an unconstrained hypothesis \[
P_u(\boldsymbol{\beta}, \phi | \boldsymbol{y}, \boldsymbol{X}) = 
\frac{
f(\boldsymbol{y} | \boldsymbol{X}, \boldsymbol{\beta}, \phi)_{1-b} ~ p_u(\boldsymbol{\beta}, \phi | \boldsymbol{y}, \boldsymbol{X}, b)
}{
\int \int f(\boldsymbol{y} | \boldsymbol{X}, \boldsymbol{\beta}, \phi)_{1-b} ~ p_u(\boldsymbol{\beta}, \phi | \boldsymbol{y}, \boldsymbol{X}, b) ~ \partial \boldsymbol{\beta} ~ \partial \phi
},
\] where
\(f(\boldsymbol{y} | \boldsymbol{X}, \boldsymbol{\beta}, \phi)_{1-b}\)
yields the fraction \(1-b\) of the likelihood.
\citet{mulder_olssoncollentine_2019} proved that updating a
\(\mathcal{T}(\boldsymbol{\hat{\beta}}, \boldsymbol{\hat{\Sigma}_\beta} / b, 1)\)
prior distribution with the remaining fraction of the likelihood of the
normal linear model yields a multivariate Student \(\mathcal{T}\)
posterior distribution \[
P_u(\boldsymbol{\beta} | \boldsymbol{y}, \boldsymbol{X}) = \mathcal{T}(\boldsymbol{\hat{\beta}}, \boldsymbol{\hat{\Sigma}_\beta}, n - k),
\] with location parameter \(\boldsymbol{\hat{\beta}}\), scale parameter
\(\boldsymbol{\hat{\Sigma}_{\beta}}\) and \(df = n - k\) degrees of
freedom.\footnote{Again, a discussion of the posterior distribution of
  the nuisance parameters is omitted, because these are integrated out
  when calculating the Bayes factors.}

For GLMs other than the normal linear regression model, as well as
hierarchical and structural equation models,
\citet{gu_approximated_2018} discussed how the posterior distribution of
the regression coefficients can be approximated by a multivariate normal
distribution, which yields \[
P_u(\boldsymbol{\beta} | \boldsymbol{y}, \boldsymbol{X}) \approx \mathcal{N}(\boldsymbol{\hat{\beta}}, \boldsymbol{\hat{\Sigma}_\beta}).
\] Because a truncated prior distribution under an informative
hypothesis yields a density that is equal to zero at the truncated part
of the distribution, and the likelihood function is unaffected by the
hypothesis, the posterior distribution under an informative hypothesis
also yields a truncated version of the unconstrained posterior. That is,
the posterior distribution under hypothesis \(H_i\) yields
\(P_i(\boldsymbol{\beta} | \boldsymbol{y}, \boldsymbol{X}) \propto P_i(\boldsymbol{\beta} | \boldsymbol{y}, \boldsymbol{X})\boldsymbol{1}_{\boldsymbol{B}_i}\).

\hypertarget{bayes-factors}{%
\subsubsection{Bayes factors}\label{bayes-factors}}

The Bayes factor comparing hypotheses \(H_i\) and \(H_{i'}\) is defined
as the ratio of the marginal likelihoods of these hypotheses, such that
\[
BF_{ii'} = \frac{m(\boldsymbol{y} | \boldsymbol{X}, H_i)}{m(\boldsymbol{y} | \boldsymbol{X}, H_{i'})}.
\] This measure can be directly interpreted as the evidence in the data
for hypothesis \(H_i\) versus the evidence in the data for hypothesis
\(H_{i'}\). As such, if \(BF_{ii'} = 7\), hypothesis \(H_i\) obtains 7
times more support than hypothesis \(H_{i'}\). The marginal likelihood
of hypothesis \(H_i\) is defined as \[
\begin{aligned}
m(\boldsymbol{y} | \boldsymbol{X}, H_i) 
&= \int \int_{\boldsymbol{\beta} \in \boldsymbol{B}_i}  P_i(\boldsymbol{\beta}, \phi | \boldsymbol{y}, \boldsymbol{X}) ~ \partial \boldsymbol{\beta} ~ \partial \phi \\
&= \int \int_{\boldsymbol{\beta} \in \boldsymbol{B}_i} f(\boldsymbol{y} | \boldsymbol{X}, \boldsymbol{\beta}, \phi) ~ p_i(\boldsymbol{\beta}, \phi | \boldsymbol{y}, \boldsymbol{X}) ~ \partial \boldsymbol{\beta} ~ \partial \phi,
\end{aligned}
\] which, when using the fractional prior and posterior, yields \[
m(\boldsymbol{y} | \boldsymbol{X}, H_i) = 
  \int \int_{\boldsymbol{\beta} \in \boldsymbol{B}_i}  
  f(\boldsymbol{y} | \boldsymbol{X}, \boldsymbol{\beta}, \phi)_{1-b} ~ 
  p_i(\boldsymbol{\beta}, \phi | \boldsymbol{y}, \boldsymbol{X}, b) ~ \partial \boldsymbol{\beta} ~ \partial \phi.
\] Whereas, in general, obtaining the marginal distribution is a
complicated endeavor, the encompassing prior approach greatly simplifies
this task \citep{klugkist_inequality_2005}. In fact,
\citet{gu_approximated_2018} showed that when comparing an informative
hypothesis \(H_i\) with an unconstrained hypothesis \(H_u\), the ratio
of the two marginal likelihoods boils down to \[
\begin{aligned}
BF_{iu} &= 
\frac{
  \int \int_{\boldsymbol{\beta} \in \boldsymbol{B}_i} f (\boldsymbol{y} | \boldsymbol{X}, \boldsymbol{\beta}, \phi)_{1-b} ~ p_i(\boldsymbol{\beta}, \phi | \boldsymbol{y}, \boldsymbol{X}, b) \partial \boldsymbol{\beta} ~ \partial \phi
}{
  \int \int f(\boldsymbol{y} | \boldsymbol{X}, \boldsymbol{\beta}, \phi)_{1-b} ~ p_u(\boldsymbol{\beta}, \phi | \boldsymbol{y}, \boldsymbol{X}, b) \partial \boldsymbol{\beta} ~ \partial \phi
} \\
&= \int \int_{\boldsymbol{\beta} \in \boldsymbol{B}_i} 
\frac{
  f (\boldsymbol{y} | \boldsymbol{X}, \boldsymbol{\beta}, \phi)_{1-b} ~  
  \frac{
    p_u (\boldsymbol{\beta}, \phi | \boldsymbol{y}, \boldsymbol{X}, b) \boldsymbol{1}_{\boldsymbol{B}_i}
  }{
    \int \int_{\boldsymbol{\beta} \in \boldsymbol{B}_i} p_u (\boldsymbol{\beta}, \phi | \boldsymbol{y}, \boldsymbol{X}, b) \partial \boldsymbol{\beta} ~ \partial \phi
  }
}{
  \int \int f(\boldsymbol{y} | \boldsymbol{X}, \boldsymbol{\beta}, \phi)_{1-b} ~ p_u(\boldsymbol{\beta}, \phi | \boldsymbol{y}, \boldsymbol{X}, b) \partial \boldsymbol{\beta} ~ \partial \phi
} \partial \boldsymbol{\beta} ~ \partial \phi \\
&= \frac{
   \int \int_{\boldsymbol{\beta} \in \boldsymbol{B}_i} P_u(\boldsymbol{\beta}, \phi| \boldsymbol{y}, \boldsymbol{X}) ~ \partial \boldsymbol{\beta} ~ \partial \phi
}{
  \int\int_{\boldsymbol{\beta} \in \boldsymbol{B}_i} p_u(\boldsymbol{\beta}, \phi | \boldsymbol{y}, \boldsymbol{X}, b) \partial \boldsymbol{\beta} ~ \partial \phi
}.
\end{aligned}
\] That is, the constrained prior can be factored into an unconstrained
prior, an indicator for the admissible parameter space under hypothesis
\(H_i\) and the normalizing marginal constrained prior distribution.
This yields a marginal prior and marginal posterior distribution,
integrated over the parameter space in line with the hypothesis.

In the current definition of the Bayes factor, however, the prior
distribution is centered at the maximum likelihood estimates of the
regression coefficients, as implied by the fraction of information from
the likelihood. This is problematic when testing inequality-constrained
hypotheses (e.g., \(H_i: \beta > 0\) versus \(H_{i'}: \beta \leq 0\)),
because one of the two hypotheses obtains a larger prior plausibility
unless the estimated coefficient lies exactly on the boundary of the
hypotheses (e.g., \(\hat{\beta}=0\)). When using a symmetric prior,
balancing the unconstrained prior on the boundary of the hypotheses
ensures that both hypotheses receive the same \emph{a priori} support.
Additionally, when testing equality-constrained hypotheses,
\citet{jeffreys_1961} remarked that centering the prior around the
hypothesized value is substantively desirable, because the \emph{a
priori} expectation is that the estimate should be close to the
hypothesized value, otherwise there is little merit in testing precisely
this hypothesis. Based on these considerations, several scholars
advocated to center the prior distribution around the focal point of
interest when calculating the Bayes factor
\citep{gu_approximated_2018, mulder_prior_2014, zellner_siow_1980, mulder_gu_bayesian_2021}.
This procedure renders the \emph{adjusted} fractional Bayes factor,
which is defined by \[
BF_{iu} = \frac{
  \int \int_{\boldsymbol{\beta} \in \boldsymbol{B}_i} P_u(\boldsymbol{\beta}, \phi | \boldsymbol{y}, \boldsymbol{X}) ~ \partial \boldsymbol{\beta} ~ \partial \phi
} {
  \int \int_{\boldsymbol{\beta} \in \boldsymbol{B}_i} p^*_u(\boldsymbol{\beta}, \phi| \boldsymbol{y}, \boldsymbol{X}, b) ~ \partial \boldsymbol{\beta} ~ \partial \phi
},
\] where the prior for the coefficients is centered around the
hypothesized values. After integrating out the nuisance parameters, this
yields \[
BF_{iu} = \frac{
  \int_{\boldsymbol{\beta} \in \boldsymbol{B}_i} P_u (\boldsymbol{\beta} | \boldsymbol{y}, \boldsymbol{X}) ~ \partial \boldsymbol{\beta}
}{
  \int_{\boldsymbol{\beta} \in \boldsymbol{B}_i} p^*_u(\boldsymbol{\beta} | \boldsymbol{y}, \boldsymbol{X}, b) ~ \partial \boldsymbol{\beta}
}
= \frac{f_i}{c_i}.
\] Consequently, \(BF_{iu}\) is defined as the proportion of the
posterior distribution that is in line with hypothesis \(H_i\), denoted
fit (\(f_i\)), over the proportion of the adjusted unconstrained prior
distribution that is in line with hypothesis \(H_i\), denoted complexity
(\(c_i\)). As such, the Bayes factor serves as Occam's razor, by
balancing the fit of the data to the hypothesis (i.e., the marginal
posterior) with the complexity of the hypothesis.

This formulation of the Bayes factor is relatively easy to compute for
both equality and inequality constrained informative hypotheses. When
the hypothesis of interest only contains equality constraints, one can
make use of the Savage-Dickey density ratio \citep{savage_dickey_1971},
which yields \[
BF_{i_0u} = \frac{f_{i_0}}{c_{i_0}} = \frac{
  P_u(\boldsymbol{\beta} = \boldsymbol{B}_{i_0} | \boldsymbol{y}, \boldsymbol{X})
}{
  p_u^*(\boldsymbol{\beta} = \boldsymbol{B}_{i_0} | \boldsymbol{y}, \boldsymbol{X}, b)
},
\] which is the ratio of the densities of the posterior
\(P_u(\boldsymbol{\beta} = \boldsymbol{B}_{i_0} | \boldsymbol{y}, \boldsymbol{X})\)
and adjusted prior
\(p^*_u(\boldsymbol{\beta} = \boldsymbol{B}_{i_0} | \boldsymbol{y}, \boldsymbol{X}, b)\),
evaluated at the location of the hypothesized values. When the
hypothesis of interest contains merely inequality constraints, the Bayes
factor is expressed as the ratio between the volume of the posterior
that is in line with the hypothesis and the volume of the prior that is
in line with the hypothesis, such that \[
BF_{i_1u} = \frac{f_{i_1}}{c_{i1}} = 
\frac{
  \int_{\boldsymbol{\beta} \in \boldsymbol{B}_{i_1}} P_u(\boldsymbol{\beta} | \boldsymbol{y}, \boldsymbol{X}) ~ \partial \boldsymbol{\beta}
}{
  \int_{\boldsymbol{\beta} \in \boldsymbol{B}_{i_1}} p_u^*(\boldsymbol{\beta} | \boldsymbol{y}, \boldsymbol{X}, b) ~ \partial \boldsymbol{\beta}
}.
\] In the most complex scenario where the hypothesis of interest
contains both equality and inequality constraints, the Bayes factor
equals \[
BF_{i_{01}u} = \frac{f_{i_{01}}}{c_{i_{01}}} = \frac{
  P_u(\boldsymbol{\beta}_{i_0} = \boldsymbol{B}_{i_0} | \boldsymbol{y}, \boldsymbol{X})
}{
  p^*_u(\boldsymbol{\beta}^*_{i_0} = \boldsymbol{B}_{i_0} | \boldsymbol{y}, \boldsymbol{X}, b)
} \times
\frac{
  \int_{\boldsymbol{\beta}_{i_1} \in \boldsymbol{B}_{i_1}} P_u(\boldsymbol{\beta}_{i_1} | \boldsymbol{\beta}_{i_0} = \boldsymbol{B}_{i_0}, \boldsymbol{y}, \boldsymbol{X}) ~ \partial \boldsymbol{\beta}
}{
  \int_{\boldsymbol{\beta}_{i_1} \in \boldsymbol{B}_{i_1}} p^*_u(\boldsymbol{\beta}_{i_1} | \boldsymbol{\beta}_{i_0} = \boldsymbol{B}_{i_0} \boldsymbol{y}, \boldsymbol{X}, b) ~ \partial \boldsymbol{\beta}
},
\] where \(\boldsymbol{\beta}_{i_0}\) and \(\boldsymbol{\beta}_{i_1}\)
are the equality and inequality constrained parameters under the
hypothesis, respectively \citep{gu_approximated_2018}. Additionally, the
Bayes factor between two informative hypotheses is easy to calculate due
to transitivity of the Bayes factor. That is, both informative
hypotheses can be expressed with reference to the unconstrained
hypothesis, such that comparing two informative hypotheses \(H_i\) and
\(H_{i'}\) yields \[
BF_{ii'} = \frac{BF_{iu}}{BF_{i'u}} = \frac{f_i / c_i}{f_{i'}/c_{i'}}.
\] A special case of an alternative informative hypothesis is a
complement hypothesis. The complement of hypothesis \(H_i\) implies
``not \(H_i\)'' and reflects all possible parameter values outside the
constraints imposed by the hypothesis under consideration. Evaluating a
single informative hypothesis against its complement yields \[
BF_{i c} = \frac{BF_{iu}}{BF_{cu}} = \frac{f_i/c_i}{(1 - f_i) / (1 - c_i)}.
\] Because a hypothesis and its complement cover mutually exclusive
regions of the entire parameter space, evaluating against the complement
is more powerful than evaluating against an unconstrained hypothesis
\citep{klugkist_volker_2023}.

When comparing a set of hypotheses, it is generally insightful to
translate the Bayes factors into posterior model probabilities
\citep[\(PMP\)s;][]{kass_raftery_bayes_factors_1995}. The posterior
model probabilities quantify the support for each hypothesis on a common
scale on the interval \((0,1)\), which facilitates the interpretation,
and allows to take prior knowledge into account through prior model
probabilities. The posterior model probabilities for hypothesis \(H_i\),
with \(i = 1, 2, \dots, m\) (with \(m\) the number of hypotheses under
consideration), are given by \[
PMP(H_{i}) = \frac{\pi_i BF_{iu}}{\sum^m_{i'=1} \pi_{i'} BF_{i'u}}, 
\] where \(\pi_i\) indicates the prior model probability of hypothesis
\(H_i\), and render the relative plausibility of a finite set of
hypotheses after observing the data.

\hypertarget{bayesian-evidence-synthesis}{%
\subsection{Bayesian Evidence
Synthesis}\label{bayesian-evidence-synthesis}}

The procedure of quantifying the support for a hypothesis using Bayes
factors can be extended to express the support for a hypothesis or
scientific theory over multiple studies \citep{kuiper_combining_2013}.
When all considered studies assess an overarching theory, the Bayes
factor in each study renders a certain degree of support for or against
this theory. Using \emph{Bayesian Evidence Synthesis} (\emph{BES}), the
aggregated support for the theory is obtained by updating the model
probabilities with each new study. In this sense, \emph{BES} proceeds
sequentially. After conducting the first study, the support for the
hypothesis of interest can be expressed as a posterior model
probability. This first posterior model probability can be used as prior
model probability for the second study. Combined with the support for
the hypothesis in the second study, this renders a posterior model
probability that reflects the total amount of evidence for or against
the hypothesis in the first and second study combined. Hence, the
posterior model probabilities after study \(t\) can be used as prior
model probabilities in study \(t + 1\) \citep{kuiper_combining_2013}.
Independent of the order of the studies, this process can be repeated
for a total of \(T\) studies, which yields \[
PMP(H_i)^{(T)} = \frac{
  \pi^{(0)}_{i} \prod^T_{t=1} BF^{(t)}_{iu}
}{
  \sum^m_{i'=1} \pi^{(0)}_{i'} \prod^T_{t=1} BF^{(t)}_{i'u}
},
\] where \(\pi^{(0)}_i\) indicates the prior model probability for
hypothesis \(H_i\) before any study has been conducted. It can be
reasonable to consider all hypotheses equally likely before observing
any data, which renders equal prior model probabilities for all
hypotheses (i.e., \(\pi^{(0)}_i = 1/m\)). With equal prior model
probabilities, the product of Bayes factors contains the same
information as the aggregated posterior model probabilities. Ultimately,
the final posterior model probabilities reflect the relative support for
the theory over all studies.

\hypertarget{simulations}{%
\section{Simulations}\label{simulations}}

Although \emph{BES} allows to combine support for a hypothesis that is
tested differently over studies, this approach also poses challenges.
Bayes factors do not only depend on the fit of the data to the
hypothesis, but also on the complexity of the hypothesis. Conceptually
similar hypotheses with distinct operationalizations in different
studies can have different complexities. Consider a hypothesis that
implies a positive relationship between a construct of interest,
measured by multiple indicators, and an outcome. Whereas some
researchers may combine multiple indicators into a single scale score,
others may analyze the effects of the separate indicators. Similarly,
researchers may categorize continuous variables. Such variation in
measurement and data handling results in hypotheses with different
complexities, affecting the resulting Bayes factors. The upcoming
section presents the first set of eight simulations. We briefly assess
how statistical power affects the performance of \emph{BES} in terms of
the aggregated support for the true hypothesis, and then zoom in on how
different choices in data handling, resulting in conceptually similar
hypotheses with different complexities, affect \emph{BES}. We first
outline the general set-up, after which we discuss the specifics per
pair of simulations. The methodological details and the results of the
respective simulation(s) are integrated to enhance the readability.
Hereafter, a second set of simulations zooms in on those scenarios in
which \emph{BES} performs inadequately. The simulations are conducted in
\texttt{R} \citep[Version 4.1.0; a full simulation archive containing
all \texttt{R}-scripts and results is available on GitHub, see Data
Availability Statement]{R}.

\hypertarget{simulations-part-1---assessing-the-effect-of-complexity}{%
\subsection{Simulations part 1 - assessing the effect of
complexity}\label{simulations-part-1---assessing-the-effect-of-complexity}}

In part one of the simulations, eight simulation conditions are
assessed. The focus of these simulations is mainly on how the complexity
of the candidate hypotheses affects the performance of \emph{BES}, while
simultaneously varying the sample and effect sizes. In all simulations,
we consistently apply \emph{BES} on a collection of three studies that
all assess the same hypothesis. The hypotheses are varied in a pairwise
manner between simulation conditions, such that a comparison is made
between conceptually similar hypotheses that have different
complexities, due to different choices in data handling. Such data
handling is performed \emph{after} generating the data (we discuss the
operationalizations of hypotheses per simulation condition in subsequent
sections). To demonstrate that \emph{BES} is applicable despite
between-study heterogeneity, data representing the three studies is
generated and analysed with three statistical models: ordinary least
squares (OLS), logistic and probit regression. Within each of the eight
simulations, we consider a common set of sample sizes
(\(n \in \{25, 50, 100, 200, 400, 800\}\)) and effect sizes
\citep[\(R^2 \in \{0.02, 0.09, 0.25\}\), corresponding to small, medium
and large effects as defined by][]{cohen_1988}, that are equivalent for
the three studies. Whereas the conventional \(R^2\) is used for studies
generated with OLS regression, McKelvey and Zavoina's \(R^2_{M\&K}\)
\citeyearpar{mckelvey_zavoina_1975} is used for studies generated with
logistic and probit models, due to its close empirical resemblance of
the conventional \(R^2\)
\citep{hagle_mitchell_goodness_1992, demaris_explained_2002}. In each
study, the Bayes factor for the hypothesis of interest is calculated
using the \texttt{R}-package \texttt{BFpack} \citep{BFpack}, after which
\emph{BES} is applied with equal initial prior model probabilities for
the hypotheses under consideration. The performance of \emph{BES} is
assessed over \(1000\) iterations for each combination of effect and
sample size. Performance is measured in terms of the aggregated Bayes
factors of the hypothesis of interest against an unconstrained
hypothesis (on a log-scale to mitigate scaling issues), and in terms of
posterior model probabilities against both an unconstrained hypothesis
and a complement hypothesis.\footnote{Note that Bayes factors against a
  complement hypothesis become infinitely large when the fit of the
  hypothesis tends to \(1\).}

In all simulation conditions, data is generated based on the same
relationship between predictors and outcome, regardless of the model
used to generate the data. Each study consists of an outcome
\(\boldsymbol{y}\), that can be continuous or dichotomous, and \(p = 6\)
predictor variables in the \(n \times p\) matrix \(\boldsymbol{X}\),
with columns
\(\boldsymbol{x}_1, \boldsymbol{x}_2, \dots, \boldsymbol{x}_6\). The
weights of the relationship between \(\boldsymbol{X}\) and
\(\boldsymbol{y}\) is captured in the column vector
\(\boldsymbol{a} = \begin{bmatrix} 0 & 1 & 1 & 1 & 2 & 3 \end{bmatrix}'\),
such that the population regression coefficients are specified as
\(\beta_1 = 0\), \(\beta_2 = \beta_3 = \beta_4\),
\(\beta_5 = 2\beta_{2,3,4}\) and \(\beta_6 = 3\beta_{2,3,4}\). All
predictor variables are normally distributed with zero mean vector
\(\boldsymbol{\mu}\) and covariance matrix \(\boldsymbol{\Sigma}\), with
common covariance \(\rho_{k,k'}=0.3\), such that \[
\boldsymbol{\mu} = 
\begin{bmatrix}
0 \\ 0 \\ 0 \\ 0 \\ 0 \\ 0
\end{bmatrix}, 
~~~~
\boldsymbol{\Sigma} = 
\begin{bmatrix}
1 &  &  &  &  &  \\ 
0.3 & 1 &  &  &  &  \\ 
0.3 & 0.3 & 1 &  &  &  \\ 
0.3 & 0.3 & 0.3 & 1 &  &  \\ 
0.3 & 0.3 & 0.3 & 0.3 & 1 &  \\ 
0.3 & 0.3 & 0.3 & 0.3 & 0.3 & 1  \\ 
\end{bmatrix}.
\] Consequently, the regression coefficients are defined by \[
\boldsymbol{\beta} = \boldsymbol{a} 
  \sqrt{
    \frac{\text{Var}(\boldsymbol{\hat{y}})}{\boldsymbol{1'}(\boldsymbol{a} \boldsymbol{a'}\odot\boldsymbol{\Sigma})\boldsymbol{1}}
  },
\] where
\(\text{Var}(\boldsymbol{\hat{y}}) = \text{Var}(\boldsymbol{X\beta})\)
is defined as a function of the effect size,\footnote{
  \(\text{Var}(\boldsymbol{\hat{y}}_{\text{OLS}}) = R^2\) under OLS when
  \(\boldsymbol{y}\) has a variance of
  \(\sigma_{\boldsymbol{y}}^2 = 1\), while
  \(\text{Var}(\boldsymbol{\hat{y}}_{\text{logistic}}) = \frac{R^2\frac{\pi^2}{3}}{1 - R^2}\)
  and
  \(\text{Var}(\boldsymbol{\hat{y}}_{\text{probit}}) = \frac{R^2}{1-R^2}\)
  for logistic and probit regression, respectively.} \(\boldsymbol{1}\)
is a \(p\)-dimensional one vector, \(\boldsymbol{1'}\) is its transpose
and \(\odot\) indicates the Hadamard (element-wise) product. The
population-level regression coefficients are displayed in Table
\ref{tbl-coefs} for all models and effect sizes (rounded at three
decimals, which may slightly distort their displayed relative sizes).
Continuous outcomes are drawn from a normal distribution \[
\boldsymbol{y}_{\text{OLS}} \sim \mathcal{N}(\boldsymbol{X\beta}, (1 - R^2)\boldsymbol{I}),
\] with mean vector \(\boldsymbol{X\beta}\), common residual variance
\(\sigma_{\epsilon}^2=1-R^2\) and \(\boldsymbol{I}\) an
\(n\)-dimensional identity matrix. Binary outcomes are drawn from a
Bernoulli distribution \[
\boldsymbol{y}_{\text{logistic}} ~ \sim ~ \mathcal{B}\Big(\frac{\exp\{\boldsymbol{X\beta}\}}{1 + \exp\{\boldsymbol{X\beta}\}}\Big), 
~ \text{and} ~~
\boldsymbol{y}_{\text{probit}} ~ \sim ~ \mathcal{B}(\Phi\{\boldsymbol{X\beta}\}),
\] with \(\Phi\) indicating the cumulative normal
distribution.\footnote{Complete separation in data with binary outcomes
  occurred sporadically for the smallest sample sizes, and was handled
  by generating a new data set.}

\begin{table}[t]
\centering
\caption{Population-level regression coefficients for ordinary least squares (OLS), logistic and probit regression, given effect sizes of $R^2 \in \{0.02, 0.09, 0.25\}$.} 
\label{tbl-coefs}
\scalebox{0.8}{
\begin{tabular}{llllllllllllllll}
  \toprule
  $R^2$ & & \multicolumn{4}{c}{OLS} & & \multicolumn{4}{c}{Logistic} & & \multicolumn{4}{c}{Probit} \\
 \midrule
  &   & $\beta_1$ & $\beta_{2, 3, 4}$ & $\beta_5$ & $\beta_6$ &   & $\beta_1$ & $\beta_{2, 3, 4}$ & $\beta_5$ & $\beta_6$ &   & $\beta_1$ & $\beta_{2, 3, 4}$ & $\beta_5$ & $\beta_6$ \\ 
   \midrule
0.02 &   & 0.000 & 0.026 & 0.051 & 0.077 &   & 0.000 & 0.047 & 0.094 & 0.141 &   & 0.000 & 0.026 & 0.052 & 0.078 \\ 
  0.09 &   & 0.000 & 0.054 & 0.109 & 0.163 &   & 0.000 & 0.103 & 0.207 & 0.310 &   & 0.000 & 0.057 & 0.114 & 0.171 \\ 
  0.25 &   & 0.000 & 0.091 & 0.181 & 0.272 &   & 0.000 & 0.190 & 0.380 & 0.570 &   & 0.000 & 0.105 & 0.209 & 0.314 \\ 
   \bottomrule
\end{tabular}
}
\end{table}

\hypertarget{simulation-1-and-2}{%
\subsubsection{Simulation 1 and 2}\label{simulation-1-and-2}}

In simulation 1 and 2, we assess the consequences of including one
underpowered study in the set of three for the performance of
\emph{BES}, while simultaneously assessing the aforementioned variations
in sample and effect size. Simulation 1 completely adheres to the
outlined set up. In simulation 2, we randomly select one of the three
studies to have a sample size of \(n = 25\). The sample sizes of the two
other studies are unaffected, and incrementally increase from \(25\) to
\(800\). Each generated study is analyzed with a regression model
containing all six predictors and an intercept (which equals 0 in the
population). We evaluate the hypothesis
\(H_{1,2}: \beta_4 < \beta_5 < \beta_6\),\footnote{ The hypotheses are
  numbered such that they correspond to the simulation in which they are
  evaluated.} and thus do not consider hypotheses with different
complexities, yet.

\begin{figure}[!t]

{\centering \includegraphics[width=1\textwidth,height=\textheight]{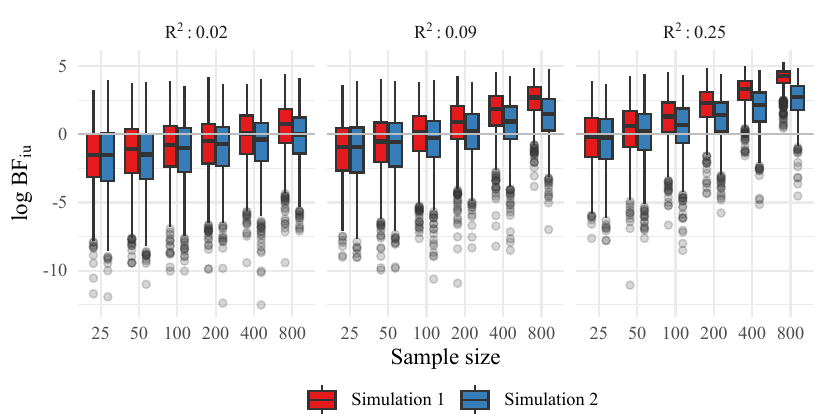}

}

\caption{\label{fig-BF12}Aggregated Bayes factors for hypothesis
\(H_{1,2}: \beta_4 < \beta_5 < \beta_6\) versus \(H_u\) over three
studies (with linear, logistic and probit models). In simulation 2, one
of the three studies is randomly selected to have a small sample size
(\(n = 25\)).}

\end{figure}

In simulation 1 and 2, the aggregated Bayes factors for the true
hypothesis \(H_{1,2}: \beta_4 < \beta_5 < \beta_6\) against the
unconstrained hypothesis \(H_u\) increase with sample size and effect
size (Figure~\ref{fig-BF12}). In simulation 1 (the red boxes), for a
small effect size (\(R^2 = 0.02\)), \(H_{1,2}\) is more often supported
than \(H_u\) only when the sample size is at least \(n \geq 400\). For
\(n = 400\) and \(R^2 = 0.02\), the median aggregated Bayes factor (the
middle horizontal black line in each box) is slightly above zero on the
logarithmic scale, indicating that there is more support for the true
hypothesis \(H_{1,2}\) than for \(H_u\) in about \(50\%\) of the
iterations. When the effect sizes increase, \(H_{1,2}\) is preferred
over \(H_u\) in most iterations when \(n \geq 100\) (when
\(R^2 = 0.09\)) or when \(n \geq 50\) (when \(R^2 = 0.25\)). For \(800\)
observations per study and a large effect, the aggregated Bayes factor
approaches an upper bound, that is due to evaluating against an
unconstrained alternative hypothesis. That is, when the hypothesis fits
the data perfectly (i.e., \(f_i = 1\)), the Bayes factor within a study
cannot exceed \(1/c_i\), and the aggregated Bayes factor cannot exceed
\(\prod_t^T 1/c_i^{(t)}\). In simulation 2 (the blue boxes), this upper
bound is not reached in any condition, as the support for \(H_{1,2}\) is
generally smaller than in simulation 1. Likewise, larger sample sizes
(for the two unaffected studies) and effect sizes are required before
\(H_{1,2}\) obtains more support than \(H_u\) in the majority of the
iterations. These findings are to be expected, because studies with the
smallest sample size often provide support against \(H_{1,2}\).
Replacing a study with a larger sample size by a study with a sample
size of \(n = 25\) thus leads to a decrease in the aggregated support.

\begin{figure}[!t]

{\centering \includegraphics[width=1\textwidth,height=\textheight]{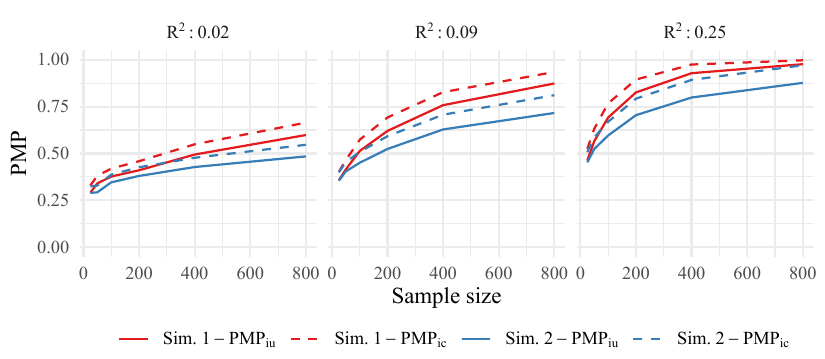}

}

\caption{\label{fig-PMP12}Aggregated \(PMP\)s for hypothesis
\(H_{1,2}: \beta_4 < \beta_5 < \beta_6\) versus \(H_u\) or \(H_c\) over
three studies (with linear, logistic and probit models). In simulation
2, one of the three studies is randomly selected to have a small sample
size (\(n = 25\)).}

\end{figure}

Additionally, the aggregated support for the true hypothesis \(H_{1,2}\)
versus both the unconstrained and complement hypothesis is quantified
with posterior model probabilities (\(PMP\)s; Figure~\ref{fig-PMP12}).
These results also show that the support for the true hypothesis
increases with the sample and effect size. For the smallest effect size,
the average aggregated support does not exceed \(0.70\), indicating that
\(H_{1,2}\) is hardly favored over \(H_u\). For medium and large
effects, the \(PMP\)s tend to \(1\) when the sample size is large
enough. Comparing simulation 1 and 2 shows that considering a single
underpowered study in the set of studies substantially reduces the
average aggregated \(PMP\)s. When the effect size equals \(R^2 = 0.09\),
the average aggregated support is larger for three studies with a sample
size of \(n = 400\) (in simulation 1; the red lines) than for two
studies with a sample size of \(n = 800\) and one study with \(n = 25\)
(in simulation 2; the blue lines). Hence, although the total number of
observations is higher in the latter setting, the average support for
the true hypothesis is not, regardless of the alternative hypothesis.
Lastly, Figure~\ref{fig-PMP12} shows that evaluating against the
complement hypothesis consistently renders more support for the true
hypothesis, although the difference is relatively small.

\hypertarget{simulation-3-and-4}{%
\subsubsection{Simulation 3 and 4}\label{simulation-3-and-4}}

Simulations 3 and 4 assess how the complexities of hypotheses affect
\emph{BES}, by varying the operationalizations of a construct of
interest. Both simulations consider the expectation that
\(\boldsymbol{x}_6\) and \(\boldsymbol{y}\) are positively related
(after controlling for \(\boldsymbol{x}_1\) to \(\boldsymbol{x}_5\)). In
simulation 3, \(\boldsymbol{x}_6\) is considered as is, and the
hypothesis \(H_3: \beta_6 > 0\) is evaluated. In simulation 4,
\(\boldsymbol{x}_6\) is categorized into equally sized tertiles in each
study, corresponding to \emph{low}, \emph{medium} and \emph{high}
scoring groups \citep[which is, despite advice against it, a common
procedure in many areas of research;
e.g.,][]{bennette_against_2012, decoster_best_2011}. Capturing the
expected positive relationship between \(\boldsymbol{x}_6\) and
\(\boldsymbol{y}\) into an informative hypothesis yields
\(H_4: \beta_{\text{low}} < \beta_{\text{medium}} < \beta_{\text{high}}\),
where each coefficient reflects the mean of that group controlled for
the five other variables. Consequently, the complexity of \(H_4\) is
substantially smaller than the complexity of \(H_3\), because more
constraints are placed on the parameters. Note, however, that
categorizing a continuous variable also reduces the information in this
variable, resulting in less statistical power.

\begin{figure}[!t]

{\centering \includegraphics[width=1\textwidth,height=\textheight]{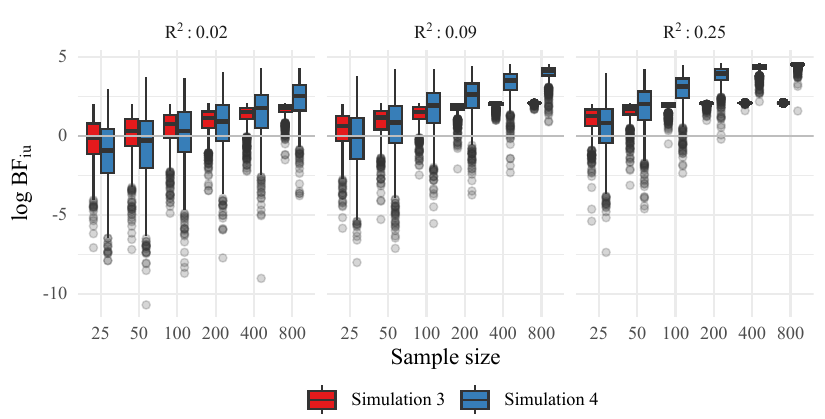}

}

\caption{\label{fig-BF34}Aggregated Bayes factors for hypothesis
\(H_3: \beta_6 > 0\) versus \(H_u\) in simulation 3 and
\(H_4: \beta_{\text{low}} < \beta_{\text{medium}} < \beta_{\text{high}}\)
versus \(H_u\) in simulation 4 over three studies (with linear, logistic
and probit models).}

\end{figure}

\begin{figure}[!t]

{\centering \includegraphics[width=1\textwidth,height=\textheight]{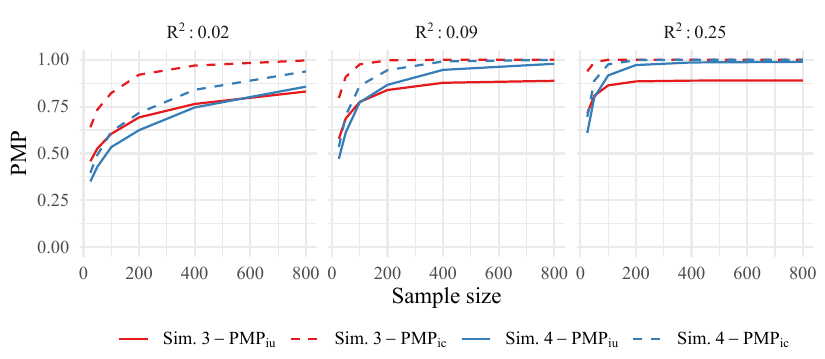}

}

\caption{\label{fig-PMP34}Aggregated \(PMP\)s for hypothesis
\(H_3: \beta_6 > 0\) versus \(H_u\) and \(H_c\) in simulation 3, and for
\(H_4: \beta_{\text{low}} < \beta_{\text{medium}} < \beta_{\text{high}}\)
versus \(H_u\) or \(H_c\) in simulation 4 over three studies (with
linear, logistic and probit models).}

\end{figure}

Figure~\ref{fig-BF34} and Figure~\ref{fig-PMP34} show that the support
for the true hypotheses \(H_3:\beta_6>0\) and
\(H_4: \beta_{\text{low}} < \beta_{\text{medium}} < \beta_{\text{high}}\)
increases with sample and effect size. This is a recurrent pattern when
the hypothesis of interest is in line with the specifications of the
parameters. In simulation 3 (the red boxes), \(H_3\) obtains more
support than \(H_u\) in more than \(50\%\) of the iterations when the
sample size is at least \(n \geq 50\) for small effect sizes. When
medium and large effect sizes are considered, all sample sizes yield
more support for \(H_3\) than \(H_u\) in the majority of the iterations.
For all effect sizes, the aggregated Bayes factor tends to its maximum
if the sample size is large enough (given a complexity of \(c_i = 1/2\)
per study in this simulation, the maximum aggregated Bayes factor is
approximately equal to \(2.08\) on a logarithmic scale). In simulation 4
(the blue boxes), there is initially more support for \(H_u\) than for
the hypothesis of interest for the smallest effect size as compared to
simulation 3. Additionally, the true hypothesis \(H_4\) obtains less
support than \(H_3\) for the smallest sample sizes considered (i.e., for
\(n \leq 100\) when \(R^2 = 0.02\), \(n \leq 50\) when \(R^2 = 0.09\)
and for \(n = 25\) when \(R^2 = 0.25\)). Both findings result from
\(H_4\) having a smaller complexity than \(H_3\), which requires more
statistical power to find support. Yet, categorizing
\(\boldsymbol{x}_6\) results in less statistical power. Whereas the
aggregated support for \(H_3\) quickly reaches its maximum, the support
for \(H_4\) continues to increase to higher levels. For \(n = 800\) when
\(R^2 = 0.09\), and for \(n \geq 400\) when \(R^2 = 0.25\), the support
for \(H_4\) also reaches its maximum, but this maximum is substantially
higher than the maximum in simulation 3, because the complexity of the
hypothesis is smaller in simulation 4.

The posterior model probabilities also clearly pinpoint the upper bound
of the support for \(H_3\) when evaluated against \(H_u\) (the red solid
line in Figure~\ref{fig-PMP34}), especially for medium and large
effects. When \(H_3\) obtains full support from the data, the posterior
model probabilities cannot exceed \(8/9\). The support for \(H_3\)
evaluated against the complement hypothesis is unrestricted (the dashed
red line), and therefore quickly tends to \(1\) for all effect sizes.
When evaluating against \(H_u\), the aggregated support for \(H_4\) (the
solid blue line in Figure~\ref{fig-PMP34}) is smaller than for \(H_3\)
for small samples but increases to almost \(1\) when the support for
\(H_3\) already reached its maximum, due to the smaller complexity of
\(H_4\). Additionally, there is only a small difference between
evaluating \(H_4\) against the unconstrained and the complement
hypothesis in terms of \(PMP\)s: the support for \(H_4\) is unconvincing
initially, but increases with the sample and effect size for both
alternatives.

\hypertarget{simulation-5-and-6}{%
\subsubsection{Simulation 5 and 6}\label{simulation-5-and-6}}

Simulations 5 and 6 also examine how the complexities of the candidate
hypotheses affect \emph{BES} by varying the operationalizations of key
constructs. Assume that the variables \(\boldsymbol{x}_2\),
\(\boldsymbol{x}_3\) and \(\boldsymbol{x}_4\) are indicators of the same
theoretical construct. In simulation 5, the three indicators are
collapsed into a single scale variable \(\boldsymbol{x}_{\text{scale}}\)
by taking the average of these variables for each observation in each of
the studies, which is common scientific practice
\citep{bauer_discrepancy_2016}. Accordingly, the hypothesis under
evaluation is \(H_5: \beta_{\text{scale}} > 0\), analysed in a model
including an intercept and \(\boldsymbol{x}_1\), \(\boldsymbol{x}_5\)
and \(\boldsymbol{x}_6\) as control variables. In simulation 6, the
variables are left as is and included separately in the analyses, such
that hypothesis \(H_6: \{\beta_2,\beta_3,\beta_4\} > 0\) is evaluated
(with the same control variables). Due to these specifications, \(H_5\)
has a larger complexity than \(H_6\), due to \(H_6\) addressing multiple
parameters. Simultaneously, taking the average of three indicators
reduces the standard error or \(\beta_{\text{scale}}\), resulting in
more power to detect an effect.

\begin{figure}[!t]

{\centering \includegraphics[width=1\textwidth,height=\textheight]{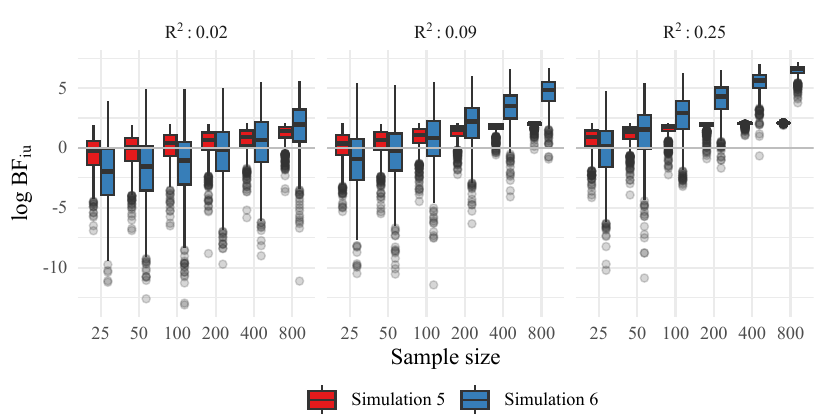}

}

\caption{\label{fig-BF56}Aggregated Bayes factors for hypothesis
\(H_5: \beta_{\text{scale}} > 0\) versus \(H_u\) in simulation 5 and
\(H_6: \{\beta_2, \beta_3, \beta_4\} > 0\) versus \(H_u\) in simulation
6 over three studies (with linear, logistic and probit models).}

\end{figure}

In simulation 5 (the red boxes in Figure~\ref{fig-BF56}), \(H_5\)
obtains more support than \(H_u\) in more than \(50\%\) of the
iterations over all sample and effect sizes, except for \(n = 25\) and
\(R^2 = 0.02\). Additionally, the aggregated support for
\(H_5: \beta_{\text{scale}}>0\) reaches its maximum for medium (at
\(n \geq 400\)) and large (at \(n \geq 100\)) effect sizes. In
simulation 6 (the blue boxes in Figure~\ref{fig-BF56}), where the
predictors are considered separately, the pattern is rather different.
When the effect size is small, the unconstrained hypothesis obtains most
support, unless the sample size is at least \(n = 400\). For medium and
large effect sizes, \(H_6\) obtains more support than \(H_u\) when
\(n \geq 100\) and \(n \geq 50\), respectively, and the support in terms
of Bayes factors tends to much higher levels than in simulation 5.

\begin{figure}[!t]

{\centering \includegraphics[width=1\textwidth,height=\textheight]{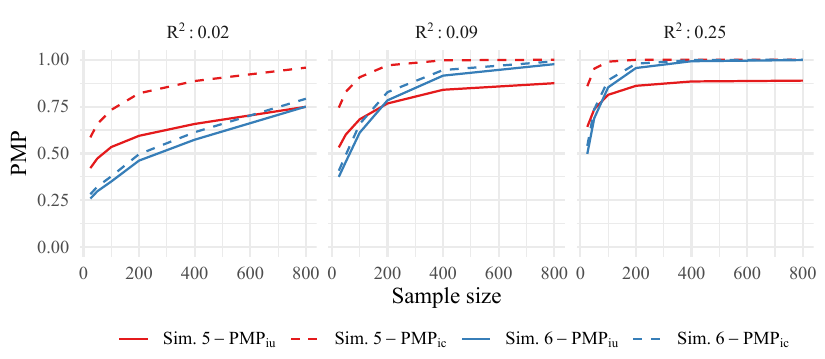}

}

\caption{\label{fig-PMP56}Aggregated \(PMP\)s for hypothesis
\(H_5: \beta_{\text{scale}} > 0\) versus \(H_u\) and \(H_c\) in
simulation 5 and \(H_6: \{\beta_2, \beta_3, \beta_4\} > 0\) versus
\(H_u\) or \(H_c\) in simulation 6 over three studies (with linear,
logistic and probit models).}

\end{figure}

Figure~\ref{fig-PMP56} for simulation 5 and 6 shows a similar pattern as
Figure~\ref{fig-PMP34} for simulation 3 and 4. The support for \(H_5\)
tends to its maximum when compared with \(H_u\) (the solid red line),
especially for a large effect, while comparing against the complement
leads to \(PMP\)s close to \(1\) (the dashed red line). In simulation 6,
the complexity of the hypothesis of interest is smaller, and the
difference between evaluating against \(H_u\) (the solid blue line) or
\(H_c\) (the dashed blue line) decreases on the scale of the \(PMP\)s.
For small sample and effect sizes, this renders small \(PMP\)s, to such
an extent that \(H_u\) is preferred over \(H_6\) for the smallest sample
and effect sizes. Yet, the smaller complexity also yields that the
maximum \(PMP\) when evaluating against \(H_u\) is larger in simulation
6 than in simulation 5. If the sample and effect size are sufficiently
large, the \(PMP\)s for \(H_6\) are close to \(1\), regardless of
whether \(H_6\) is evaluated against \(H_u\) or \(H_c\). These results
show that more statistical power is required in order to evaluate more
specific hypotheses.

\hypertarget{simulation-7-and-8}{%
\subsubsection{Simulation 7 and 8}\label{simulation-7-and-8}}

All previous simulations were concerned about whether \emph{BES}
provides adequate results when a correct informative hypothesis is
evaluated against the unconstrained or complement hypothesis. In
simulation 7 and 8, we assess the performance of \emph{BES} when the
hypothesis of interest is incorrect. We therefore expect \emph{BES} to
render \emph{less} support for the hypotheses of interest when the
sample and effect size increase. In simulation 7, we consider
\(H_7: \{\beta_2, \beta_3, \beta_4\} < 0\), implying a negative
relationship between the indicators \(\boldsymbol{x}_2\),
\(\boldsymbol{x}_3\) and \(\boldsymbol{x}_4\) and the outcome
\(\boldsymbol{y}\), which is for each parameter in the opposite
direction of the true value. Hence, the unconstrained and complement
alternative hypotheses are correct, although rather unspecific, in these
simulations. In simulation 8, we evaluate the \emph{partially incorrect}
hypothesis \(H_8: \{\beta_1, \beta_2, \beta_3\} > 0\), which is correct
for \(\beta_2\) and \(\beta_3\), but incorrect for \(\beta_1\), because
\(\beta_1 = 0\). This specification renders the unconstrained hypothesis
correct, while the complement hypothesis is also partially incorrect,
because the true parameter value of \(\beta_1\) is exactly on the
boundary of \(H_8\) and \(H_c\). In both simulations, the analysis model
contains an intercept and the remaining variables as control variables.

\begin{figure}[!t]

{\centering \includegraphics[width=1\textwidth,height=\textheight]{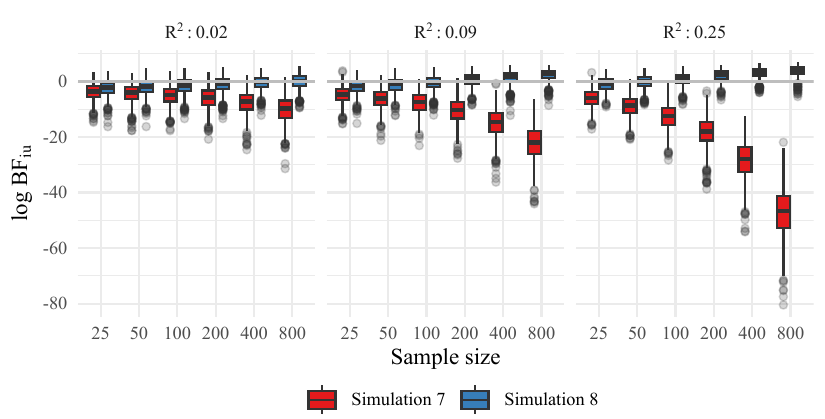}

}

\caption{\label{fig-BF78}Aggregated Bayes factors for \textit{incorrect}
and \textit{partially incorrect} hypotheses
\(H_7: \{\beta_2, \beta_3, \beta_4\} < 0\) and
\(H_8: \{\beta_1, \beta_2, \beta_3\} > 0\) versus \(H_u\) in simulation
7 and 8 over three studies (with linear, logistic and probit models).}

\end{figure}

\begin{figure}[!t]

{\centering \includegraphics[width=1\textwidth,height=\textheight]{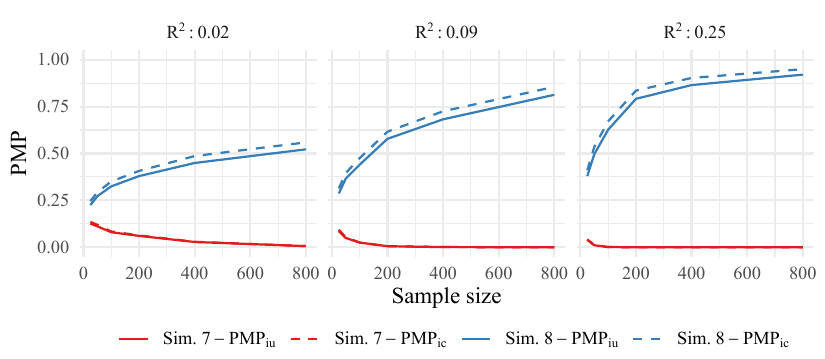}

}

\caption{\label{fig-PMP78}Aggregated \(PMP\)s for \textit{incorrect} and
\textit{partially incorrect} hypotheses
\(H_7: \{\beta_2, \beta_3, \beta_4\} < 0\) and
\(H_8: \{\beta_1, \beta_2, \beta_3\} > 0\) versus \(H_u\) or \(H_c\) in
simulation 7 and 8 over three studies (with linear, logistic and probit
models). Note that the lines of \(PMP_{i,u}\) and \(PMP_{i,c}\) are
almost completely overlapping in simulation 7.}

\end{figure}

The red boxes in Figure~\ref{fig-BF78} show that in simulation 7, the
support for the incorrect hypothesis of interest quickly decreases,
rendering more support for the unconstrained hypothesis. In fact,
already from the smallest sample sizes onward, there is less support for
\(H_7\) than for the correct hypothesis \(H_u\), which further decreases
when the sample and effect size increase. In simulation 8, the support
for the partially incorrect hypothesis \(H_8\) increases, rather than
decreases, with the sample size and effect size (the blue boxes in
Figure~\ref{fig-BF78}). Whereas for small sample and effect sizes the
correct unconstrained hypothesis is preferred over \(H_8\), the
partially incorrect \(H_8\) eventually obtains more support. Although
this behavior is undesirable, it is not surprising. The posterior
distribution of \(\beta_1\) is, on average, for \(50\%\) in line with
the constraints of \(H_8\), because the true parameter value is on the
boundary of the hypothesis, while for larger sample and effect sizes,
the posterior of \(\beta_2\) and \(\beta_3\) is almost completely in
line with this hypothesis. Hence, even though the hypothesis is
partially incorrect, the fit generally exceeds the complexity.

Figure~\ref{fig-PMP78} tells a similar story. The average aggregated
\(PMP\)s for the incorrect \(H_7\) render very strong support against
this hypothesis, for all sample sizes and effect sizes and for both
alternative hypotheses (the red solid and dashed lines, note that these
are almost completely overlapping). In simulation 8 (the blue solid and
dashed lines in Figure~\ref{fig-PMP78}), the average aggregated \(PMP\)s
are indecisive under a small effect size, but favor the partially
incorrect hypothesis \(H_8\) over the correct unconstrained and
partially incorrect complement hypothesis when the effect and sample
size increase. The difference between evaluating against \(H_u\) and
\(H_c\) is negligible, although \(H_u\) is the only correct hypothesis
in this simulation.

\hypertarget{discussion-simulations-part-1}{%
\subsubsection{Discussion simulations part
1}\label{discussion-simulations-part-1}}

The first simulations show that \emph{BES} performs adequately, in the
sense that the aggregated support for correct hypotheses increases with
the sample and effect size (simulations 1-6). Moreover, simulation 7
shows that incorrect hypotheses obtain less support when the sample and
effect size increase. The main focus of these simulations was on how the
complexity of the hypotheses affects the performance of \emph{BES}.
Especially in the context of conceptual replications, researchers might
want to aggregate the support for hypotheses with different
complexities, due to different operationalizations or measurement
instruments. Unlike conventional research synthesis approaches,
\emph{BES} allows to aggregate support for equivalent hypotheses with
different complexities and, as simulations 3 to 6 show, renders
satisfactory results when the individual studies have sufficient power.
Hence, the support for an overarching theory over studies can be
quantified with \emph{BES}, even when studies use different
operationalizations. Our findings may even insinuate that, with
sufficient power, common procedures in data handling that result in
hypotheses with smaller complexities, such as categorizing continuous
variables or assessing individual indicators rather than scale scores,
can lead to more support on the aggregate level. After all, the
hypotheses with the smallest complexities within a set of studies have
the highest upper bound for the Bayes factor against the unconstrained
hypothesis. However, this conclusion should be drawn with caution, if at
all, because it only holds when evaluating against the unconstrained,
but not when evaluating against the complement hypothesis. The Bayes
factor against the complement tends to infinity when the data fits the
hypothesis perfectly, and thus does not benefit from evaluating more
specific hypotheses.

In simulation 8, \emph{BES} functions unsatisfactorily, because with
increasing sample and effect sizes, it renders increasing support for
the partially incorrect hypothesis. However, this hypothesis is very
close to the truth \citep[Bayes factors have been proven to prefer the
model that is closest, in terms of Kullback-Leibler divergence, to the
true data-generating model; see][]{ly_bf_2016, berger2013statistical}.
When evaluating inequality-constrained hypotheses against the
unconstrained or complement alternative, support for incorrect
hypotheses can only keep increasing with the sample and effect size when
no parameters truly violate the hypothesis of interest (i.e, all
parameters are on or within the boundary values of the hypothesis). If
at least one of the parameters falls outside the constraints imposed by
the hypothesis, the fit will tend to zero for sufficiently large sample
sizes. Yet, the fact that partially incorrect hypotheses can obtain
substantial support warrants that researchers not only assess the
aggregated support for complex hypotheses, but also consider the support
for the parameters in the individual studies. If some of the estimated
parameters are typically in line with the hypothesis while others
fluctuate around the hypothesized value between studies, further
investigation is required, potentially leading to theoretical
refinements, that should be evaluated with new data, regardless of the
aggregated support.

Similarly to previous work \citep[e.g.,][]{klugkist_volker_2023}, our
simulations confirm that if individual studies lack statistical power,
\emph{BES} has difficulties finding support for the correct hypothesis.
Specifically, including a single underpowered study, as in simulation 1
and 2, reduces the performance of \emph{BES}. Moreover, aggregating over
three adequately powered studies provides more support for the correct
hypothesis than aggregating over two adequately powered studies and a
single underpowered study, even if the total number of observations is
higher in the latter scenario. Whereas conventional approaches for
research synthesis as meta-analysis or Bayesian sequential updating can
overcome such power issues, \emph{BES} typically does not, because it
answers a different question. Whereas these conventional approaches
assess whether the pooled estimate is in line with the hypothesis,
\emph{BES} questions to what extent each study provides support for the
hypothesis and combines this into a single measure of evidence. Hence,
some support against the hypothesis of interest in each study will
accumulate when aggregating over studies, and considerable support
against this hypothesis in one study is not necessarily counterbalanced
by support for this hypothesis in multiple other studies.

Such power issues are amplified under hypotheses with smaller
complexities. With insufficient power, sampling variability increases
and parameters may not be estimated accurately. If a hypothesis places
constraints on multiple parameters, it becomes increasingly likely that
at least one constraint is violated, reducing the fit of the hypothesis.
Common procedures in data handling that reduce the complexity of the
hypothesis (e.g., categorizing continuous variables), and simultaneously
the power of the analysis, may thus have adverse consequences for
\emph{BES}. Although our previous simulations show the vulnerability of
\emph{BES} to power issues, substantial uncertainty remains about the
extent to which these problems depend on (i) the number of studies
included, (ii) the complexity of the hypothesis of interest, and (iii)
the choice of alternative hypothesis. Evaluating against the complement
hypothesis yields a more powerful evaluation of the hypothesis of
interest, which may reduce the severity of power issues. Our findings
also suggest that including more studies with insufficient power may
lead to decreasing support for the correct hypothesis. We address these
questions in the subsequent section.

\hypertarget{simulations-part-2---power-issues-of-bes}{%
\subsection{\texorpdfstring{Simulations part 2 - power issues of
\emph{BES}}{Simulations part 2 - power issues of BES}}\label{simulations-part-2---power-issues-of-bes}}

Part two of the simulations zooms in on the extent to which \emph{BES}
renders support for \emph{correct} hypotheses as the number of studies
increases, while comparing between studies with and without sufficient
power. Moreover, we further assess the influence of the complexity of
the hypothesis and the choice of alternative hypothesis (i.e., an
unconstrained or a complement alternative hypothesis) on the amount of
aggregated support. We consider the same data-generating mechanism as in
previous simulations, but restrict the simulations to OLS regression for
the sake of brevity (although other models yield equivalent results). To
keep the results tractable, only one effect size and two sample sizes
are considered (\(R^2 = 0.09\) and \(n \in \{25, 200\}\)), representing
underpowered and adequately powered studies, respectively. The number of
studies is varied from \(1\) to \(150\), and the number of iterations is
again set to 1000. In all three simulations in part two, the cumulative
aggregated Bayes factor is assessed for the hypotheses of interest
against an unconstrained and a complement alternative hypothesis for the
two sample sizes considered.

\hypertarget{simulations-9-10-and-11}{%
\subsubsection{Simulations 9, 10 and 11}\label{simulations-9-10-and-11}}

\begin{figure}[!h]

{\centering \includegraphics[width=1\textwidth,height=\textheight]{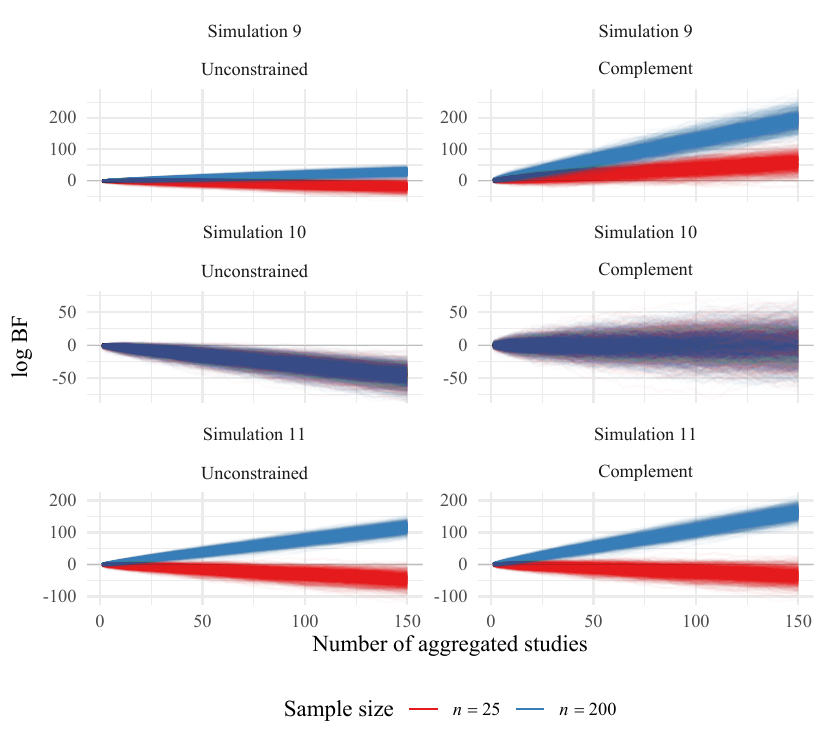}

}

\caption{\label{fig-BF91011}Cumulative aggregated Bayes factors for the
correct hypothesis \(H_9: \beta_2 > 0\) in simulation 9,
\(H_{10}: \beta_1 > 0\) in simulation 10 and
\(H_{11}: \{\beta_2, \beta_3, \beta_4\} > 0\) in simulation 11, versus
an unconstrained and complement alternative hypothesis over 1 to 150
studies, based on OLS regression models. Note that (i) the scale of the
y-axis differs between the simulations, and (ii) the lines for the two
sample sizes are almost completely overlapping in simulation 10.}

\end{figure}

In simulation 9, we evaluate \(H_9: \beta_2 > 0\) against the
unconstrained and complement hypotheses and assess the influence of the
alternative hypothesis for underpowered and adequately powered studies.
Recall that \(H_9\) is correct, as the population value of the
coefficient equals \(\beta_2 = 0.054\) (Table \ref{tbl-coefs}). For
studies with a small sample size, Figure~\ref{fig-BF91011} shows that
evaluating against \(H_u\) renders decreasing support for the correct
hypothesis \(H_9\) when the number of studies increases. After
conducting \(150\) studies, \(H_u\) is clearly preferred over \(H_9\).
For larger samples (i.e., \(n = 200\)), the correct hypothesis \(H_9\)
is consistently preferred over \(H_u\). Evaluating \(H_9\) against the
complement hypothesis consistently renders support for \(H_9\) for both
sample sizes, although the support increases faster for larger samples.
Additionally, the absence of an upper bound yields that the support for
\(H_9\) increases at a faster rate when evaluated against \(H_c\)
instead of against \(H_u\).

In simulation 10, we investigate a similar, but incorrect, hypothesis
\(H_{10}: \beta_1 > 0\) (\(\beta_{1}=0\) in the population; Table
\ref{tbl-coefs}). This renders \(H_u\) the only true hypothesis, because
\(H_c\) and \(H_{10}\) are equally incorrect. Evaluating against the
unconstrained hypothesis renders more support for \(H_u\) than for
\(H_{10}\), regardless of the sample size. When the true parameter value
is on the boundary of the hypothesis of interest, comparing against the
complement renders substantial variability in the support for \(H_{10}\)
and \(H_c\), regardless of the sample size (note that the lines for the
two sample sizes are almost completely overlapping). Although \(H_{10}\)
and \(H_c\) obtain, on average, the same support, the individual
iterations show considerable support for either of the two. Hence, there
is a downside to evaluating against the complement hypothesis, because
it is possible to find considerable support for or against a hypothesis
on the aggregate level, while there is no effect in reality.

Simulation 11 evaluates the \emph{correct}
\(H_{11}: \{\beta_2, \beta_3, \beta_4\} > 0\)
(\(\beta_2 = \beta_3 = \beta_4 = 0.054\) in the population; Table
\ref{tbl-coefs}). Whereas the hypotheses of interest in simulations 9
and 10 were balanced with their respective complements (i.e., both have
a complexity of \(c_i = 1/2\)), \(H_{11}\) is not. Also note that the
size of the coefficients here is the same as the size of the coefficient
in simulation 9, so the only difference is the complexity of the
hypothesis due to considering three parameters. Simulation 11 in
Figure~\ref{fig-BF91011} shows that the support for the correct
hypothesis \(H_{11}\) decreases when more small sample studies are
included, regardless of the alternative hypothesis, although there is
somewhat more support on average when evaluating against the complement.
Hence, when the complexity of the hypothesis of interest and its
complement are not balanced, the aggregated Bayes factor does not
necessarily tend to the \emph{correct} hypothesis, but favors the more
general alternatives. When the sample size is sufficiently large, the
issue dissolves, and the support for \(H_{11}\) increases substantially
when more studies are added, for both alternative hypotheses.

Informative hypotheses on multiple parameters can be separated in
multiple, simpler hypotheses. For example, the hypothesis \(H_{11}\) can
be deconstructed into three simpler hypotheses such that each component
is balanced with its complement (\(H_{11_a}: \beta_2 > 0\),
\(H_{11_b}: \beta_3 > 0\) and \(H_{11_c}: \beta_4 > 0\)). Evaluating
each sub-hypothesis against the unconstrained hypothesis and its
complements yields an identical outcome as in simulation 9. Each
sub-hypothesis obtains more support than the unconstrained hypothesis
when the sample size is large, but \(H_u\) obtains more support for
small sample sizes. When evaluating each hypothesis against its
complement, the sub-hypotheses obtain considerable support, regardless
of the sample size. Thus, whereas evaluating a multifaceted hypothesis
as \(H_{11}\) leads to support against the true hypothesis of interest
for smaller sample sizes, evaluating the separate components of this
hypothesis provides support for each component.

\hypertarget{discussion-simulation-part-2}{%
\subsubsection{Discussion simulation part
2}\label{discussion-simulation-part-2}}

Part two of the simulations further assessed the sensitivity of
\emph{BES} to power issues. When the studies are adequately powered, the
support for the true hypothesis is consistently found over the
situations. However, when evaluating a correct hypothesis on a single
regression parameter in studies that lack statistical power, the
aggregated support for the unconstrained hypothesis increases with the
number of studies. If the hypothesis of interest is balanced with its
complement, such that both have the same complexity, evaluating against
the complement hypothesis is not sensitive to the power within the
studies. This difference is due to the fact that evaluating against the
unconstrained hypothesis weighs evidence against the hypothesis of
interest heavier than evidence for this hypothesis. That is, the Bayes
factor for a hypothesis evaluated against the unconstrained has an upper
bound, while the lower bound of \(0\) corresponds to infinite support
for the unconstrained. Note that when studies have little power,
parameter estimates vary considerably. If a single study finds, by
chance, substantial support against the hypothesis of interest (e.g., a
fit of \(f_i=0.1\) and a complexity of \(c_i=0.5\), rendering a Bayes
factor of \(BF_{iu}=0.2\)), adding two new studies that perfectly fit
the hypothesis of interest (resulting in a Bayes factor of
\(BF_{iu}=1/0.5=2\) per study) yields an aggregate Bayes factor of
\(BF^T_{iu}=0.8\). On the aggregate level, there thus is still more
support for the unconstrained hypothesis. Evaluating against the
complement weighs the evidence for both hypotheses equally heavy under
these circumstances, and thus provides increasing support for the
correct hypothesis.

When the hypothesis of interest places constraints on multiple
parameters, evaluating against the complement results in somewhat more
power than evaluating against the unconstrained, but both provide
support against the correct hypothesis of interest when the studies are
underpowered. Whereas part one already showed that evaluating hypotheses
with smaller complexities is only appropriate when studies are
adequately powered, part two of the simulations shows that adding more
studies provides no solution. The hypothesis of interest constrained
three parameters, resulting in 8 possible parameter orderings of which
only one was correct. When the individual studies have little power, the
probability that at least one of the estimated parameters falls outside
the constraints imposed by the hypothesis is considerable, due to the
fact that the parameters are generally estimated inaccurately.
Regardless of which constraint is violated, the hypothesis may fit
poorly in each study, resulting in aggregated support against the
hypothesis that further increases with the number of studies. This issue
can be remedied by deconstructing the hypothesis space in smaller
regions, and evaluating each part separately against each complement.
This approach has two advantages. First, less statistical power is
required to find support for the correct hypothesis. Especially, if each
sub-hypothesis is balanced with its complement, \emph{BES} will provide
support for the correct hypothesis if enough studies are included.
Second, this approach allows to evaluate whether each component of the
hypothesis obtains support after aggregation. If the overall hypothesis
is incorrect, evaluating the sub-hypotheses highlights which parts of
the overall hypothesis are not supported. In this sense, deconstructing
the hypothesis space can help to further refine the theory.

Although evaluating hypotheses against their balanced complements has
advantages, it is no panacea. Simulation 10 shows that if the true
parameter value is on the boundary of the hypothesis of interest and its
complement, it is possible to find overwhelming support for either of
the two. The implications hereof reach beyond \emph{BES}, because
already within individual studies strong support can be obtained for
incorrect hypotheses. \emph{BES} further amplifies this problem. The
variability of the aggregated support increases with the number of
studies, such that overwhelming support for one of the two hypotheses
regularly occurs. This problem can be dealt with in three ways. First,
one could consider an equality-constrained hypothesis (e.g.,
\(H_i: \beta_1=0\)). In contrast to evaluating inequality-constrained
hypotheses \citep{klugkist_bf_2007}, however, evaluating
equality-constrained hypotheses with a Bayes factor is sensitive to the
scale (i.e., variance) of the prior distribution within a study
\citep{hoijtink_prior_2021, tendeiro_kiers_2019}. Accordingly, a
sensitivity analysis of the Bayes factor within a study is generally
required \citep{hoijtink_prior_2021}, which induces uncertainty
regarding the ``correct'' Bayes factor within a study, let alone when
aggregated over studies. Additionally, Bayes factors on
equality-constrained hypotheses versus unconstrained alternatives tend
to provide overly strong evidence in favor of the former, especially
when the power of the study is relatively small
\citep[e.g.,][]{tendeiro_kiers_2019}. This approach might induce more
severe power issues when using \emph{BES}. Future research should
address these considerations in more detail. Second, one could evaluate
a hypothesis with a boundary on the smallest effect of interest
\citep{lakens2014performing} versus its complement. Ultimately, when
support for the hypothesis is found, one can be sure that there is an
effect that is relevant. If no support is found, either there is no
effect, or the effect is too small to be relevant. Third, the hypothesis
of interest can be evaluated against both the unconstrained and the
complement hypothesis. If both render support for (or against) the
hypothesis of interest, one can conclude that this hypothesis provides
an accurate description of the data. If the results are contradictory,
researchers may have to acknowledge that considerable uncertainty
remains, and that more (adequately powered) studies on the topic are
required for a robust conclusion.

\hypertarget{conclusion}{%
\section{Conclusion}\label{conclusion}}

In multiple simulations, we examined the performance of Bayesian
Evidence Synthesis when evaluating hypotheses with different
complexities against different alternatives under various sample and
effect sizes. Part one of the simulations showed that \emph{BES} is
applicable regardless of differences in analysis models in the set of
studies under consideration, and renders satisfactory results if the
individual studies have sufficient statistical power. The simulations
emphasized the importance of power when aggregating evidence over
studies, especially when evaluating more specific hypotheses. For small
sample and effect sizes, evaluating a hypothesis with a relatively small
complexity generally results in support for the unconstrained and
complement hypotheses. If the Bayes factors within the studies yield
more support for the alternatives than for the specific correct
hypothesis, adding more studies that are also underpowered will not
solve the issue. Hence, \emph{BES} differs from conventional research
synthesis methods as (Bayesian) meta-analysis and Bayesian sequential
updating. Whereas the former two approaches increase the statistical
power when incorporating evidence from additional studies, \emph{BES},
in general, does not. As said, \emph{BES} answers a somewhat different
question than these conventional methods, as it aggregates the extent to
which each study supports the overall hypothesis, rather than to what
extend the pooled parameter estimate supports the hypothesis of
interest. In this sense, \emph{BES} can be seen as a joint Bayes factor
that measures the support for an overall hypothesis that states that
there is support for each study-specific hypothesis.

Part two of the simulations underscored that adding more underpowered
studies only amplifies power issues. Aggregating the evidence for a
hypothesis with a small complexity consistently renders support against
this hypothesis when the studies lack power, regardless of the
alternative hypothesis. If the hypothesis of interest and its complement
are balanced, however, the aggregated Bayes factor will eventually show
support for the correct hypothesis, if either the hypothesis of interest
or its complement is correct. As a consequence, it can be worthwhile to
separate a specific hypothesis with multiple parameter constraints into
multiple sub-hypotheses that are all balanced with their complements. If
either the sub-hypothesis or its complement is correct, the aggregated
Bayes factor will provide support for the correct hypothesis. Moreover,
evaluating sub-hypotheses allows to assess which constraints imposed by
the overall hypothesis are supported by the data and which are not,
providing a more detailed overview of the support in the studies for the
overarching hypothesis.

Overall, evaluating against the complement has advantages over
evaluating against the unconstrained. Evaluating against the complement
results in more statistical power, and the resulting Bayes factor has no
upper bound. Additionally, in a practical research setting, some studies
may have insufficient power, while others have sufficient power. In such
instances, adequately powered studies will eventually provide decisive
support when evaluated against the complement hypothesis. When
evaluating against the unconstrained, the underpowered studies may
jeopardize the aggregated evidence. That is, if underpowered studies
occasionally render support against the hypothesis of interest, the
adequately powered studies may not be able to compensate because of the
upper bound. A disadvantage of evaluating against the complement occurs
if the true parameter value is on the boundary of the hypothesis of
interest and its complement, because the aggregated support becomes
highly variable. We discussed several approaches to deal with this
issue, but future research should compare the advantages and
disadvantages of these approaches.

There are other aspects that this research did not address. First of
all, we focused solely on inequality constrained hypotheses, whereas
equality constrained hypotheses (such as the classical null hypothesis)
are still standard practice in research. As said, evaluating null
hypotheses may amplify power issues and thus affect the performance of
\emph{BES}, therefore warranting future research. Additionally, we
considered only three members of the GLM family and used relatively
simple data generating models with few parameters. When considering many
parameters, each individual parameter may only contribute little when
predicting the outcome, which would reduce the power of the analyses,
enlarging the issues we already discussed. Lastly, we varied the
hypothesis of interest only between the set of simulations, but not
within a set of studies, to easily compare between different hypotheses.
In practice, it seems more realistic that researchers want to use
\emph{BES} especially because specifications of the overall hypothesis
differ within the set of studies. Under these circumstances, the
aggregated support may be driven by results in one or two studies.
Future research should address to what extent this behavior is
problematic, and methodological developments might be required to make
\emph{BES} less sensitive to the results of ``outlying'' studies.
Potentially, weighting the studies in terms of power might be one way
forward. To some extent, this already happens implicitly when evaluating
against the complement hypothesis by allowing the Bayes factor to tend
to 0 or infinity, but not so much when evaluating against the
unconstrained hypothesis.

Overall, \emph{BES} has strengths that conventional methods for research
synthesis lack, in the sense that \emph{BES} is capable of aggregating
the support for hypotheses over studies with different designs (i.e.,
experimental, cross-sectional or longitudinal) or different
methodologies. However, there are situations in which solely evaluating
the aggregated evidence can fall short, for example when the hypothesis
of interest is too specific for the statistical power of a study, or
when the hypothesis under evaluation is partially incorrect. Hence,
researchers should not only blindly follow the results after
aggregation, but also assess the results in the individual studies. The
results on the level of the individual studies may give an additional
sense of the robustness of the results over different situations, while
simultaneously signifying potential moderating circumstances of the
effect of interest. Such additional information might raise doubts about
the robustness of the conclusions in specific scenarios, but can also
hint towards interesting new areas of research or corroborate the
conclusions from the synthesis. Overall, \emph{BES} provides great
opportunities to aggregate scientific evidence over heterogeneous
studies, but solely relying on this aggregate derogates from the wealth
of information in the individual studies.

\hypertarget{acknowledgements}{%
\section{Acknowledgements}\label{acknowledgements}}

We gratefully acknowledge stimulating discussions with Vincent Buskens
and Werner Raub.

\renewcommand\refname{References}
  \bibliography{thesis.bib}

\begin{thebibliography}{68}
\providecommand{\natexlab}[1]{#1}
\providecommand{\url}[1]{\texttt{#1}}
\expandafter\ifx\csname urlstyle\endcsname\relax
  \providecommand{\doi}[1]{doi: #1}\else
  \providecommand{\doi}{doi: \begingroup \urlstyle{rm}\Url}\fi

\bibitem[Asendorpf et~al.(2016)Asendorpf, Conner, de~Fruyt, De~Houwer, Denissen, Fiedler, Fiedler, Funder, Kliegl, Nosek, Perugini, Roberts, Schmitt, van Aken, Weber, and Wicherts]{asendorpf_recommendations_2016}
Jens~B. Asendorpf, Mark Conner, Filip de~Fruyt, Jan De~Houwer, Jaap J.~A. Denissen, Klaus Fiedler, Susann Fiedler, David~C. Funder, Reinhold Kliegl, Brian~A. Nosek, Marco Perugini, Brent~W. Roberts, Manfred Schmitt, Marcel A.~G. van Aken, Hannelore Weber, and Jelte~M. Wicherts.
\newblock \emph{Recommendations for increasing replicability in psychology.}
\newblock Methodological issues and strategies in clinical research, 4th ed. American Psychological Association, 2016.
\newblock \doi{10.1037/14805-038}.

\bibitem[Baker(2016)]{baker_reproducibility_2016}
Monya Baker.
\newblock Reproducibility crisis.
\newblock \emph{Nature}, 533\penalty0 (26):\penalty0 353--66, 2016.
\newblock \doi{10.1038/533452a}.

\bibitem[Bauer and Curran(2016)]{bauer_discrepancy_2016}
Daniel~J. Bauer and Patrick~J. Curran.
\newblock \emph{The discrepancy between measurement and modeling in longitudinal data analysis.}, pages 3--38.
\newblock CILVR series on latent variable methodology. IAP Information Age Publishing, Charlotte, NC, US, 2016.

\bibitem[Bennette and Vickers(2012)]{bennette_against_2012}
Caroline Bennette and Andrew Vickers.
\newblock Against quantiles: categorization of continuous variables in epidemiologic research, and its discontents.
\newblock \emph{BMC Medical Research Methodology}, 12\penalty0 (1):\penalty0 21, 2012.
\newblock \doi{10.1186/1471-2288-12-21}.

\bibitem[Berger(2013)]{berger2013statistical}
James~O Berger.
\newblock \emph{Statistical decision theory and Bayesian analysis}.
\newblock Springer Science \& Business Media, New York, NY, 2013.

\bibitem[Brandt et~al.(2014)Brandt, IJzerman, Dijksterhuis, Farach, Geller, Giner-Sorolla, Grange, Perugini, Spies, and Van't~Veer]{brandt_et_al_replication_2014}
Mark~J Brandt, Hans IJzerman, Ap~Dijksterhuis, Frank~J Farach, Jason Geller, Roger Giner-Sorolla, James~A Grange, Marco Perugini, Jeffrey~R Spies, and Anna Van't~Veer.
\newblock The replication recipe: What makes for a convincing replication?
\newblock \emph{Journal of Experimental Social Psychology}, 50:\penalty0 217--224, 2014.
\newblock \doi{10.1016/j.jesp.2013.10.005}.

\bibitem[Camerer et~al.(2016)Camerer, Dreber, Forsell, Ho, Huber, Johannesson, Kirchler, Almenberg, Altmejd, Chan, et~al.]{camerer2016evaluating}
Colin~F Camerer, Anna Dreber, Eskil Forsell, Teck-Hua Ho, J{\"u}rgen Huber, Magnus Johannesson, Michael Kirchler, Johan Almenberg, Adam Altmejd, Taizan Chan, et~al.
\newblock Evaluating replicability of laboratory experiments in economics.
\newblock \emph{Science}, 351\penalty0 (6280):\penalty0 1433--1436, 2016.
\newblock \doi{10.1126/science.aaf0918}.

\bibitem[Camerer et~al.(2018)Camerer, Dreber, Holzmeister, Ho, Huber, Johannesson, Kirchler, Nave, Nosek, Pfeiffer, et~al.]{camerer2018evaluating}
Colin~F Camerer, Anna Dreber, Felix Holzmeister, Teck-Hua Ho, J{\"u}rgen Huber, Magnus Johannesson, Michael Kirchler, Gideon Nave, Brian~A Nosek, Thomas Pfeiffer, et~al.
\newblock Evaluating the replicability of social science experiments in nature and science between 2010 and 2015.
\newblock \emph{Nature Human Behaviour}, 2\penalty0 (9):\penalty0 637--644, 2018.
\newblock \doi{10.1038/s41562-018-0399-z}.

\bibitem[Cohen(1988)]{cohen_1988}
J.~Cohen.
\newblock \emph{Statistical power analysis for the behavioral sciences}.
\newblock Lawrence Erlbaum Associates, New York, NY, 2nd edition, 1988.

\bibitem[Cohen(1990)]{cohen_things_i_learned_1990}
Jacob Cohen.
\newblock Things {I} have learned (so far).
\newblock \emph{American Psychologist}, 45\penalty0 (12):\penalty0 1304--1312, 1990.
\newblock \doi{10.1037/0003-066X.45.12.1304}.

\bibitem[Cohen(1994)]{cohen_earth_1994}
Jacob Cohen.
\newblock The earth is round (p{\enspace}<{\enspace}.05).
\newblock \emph{American Psychologist}, 49\penalty0 (12):\penalty0 997--1003, 1994.
\newblock \doi{10.1037/0003-066X.49.12.997}.

\bibitem[Cooper et~al.(2009)Cooper, Hedges, and Valentine]{cooper_handbook_2009}
Harris Cooper, Larry~Vernon Hedges, and Jeffrey~C Valentine.
\newblock \emph{The Handbook of Research Synthesis and Meta-Analysis}.
\newblock Russell Sage Foundation, New York, NY, 2nd edition, 2009.

\bibitem[Crandall and Sherman(2016)]{crandall_conceptual_2016}
Christian~S. Crandall and Jeffrey~W. Sherman.
\newblock On the scientific superiority of conceptual replications for scientific progress.
\newblock \emph{Journal of Experimental Social Psychology}, 66:\penalty0 93--99, 2016.
\newblock \doi{10.1016/j.jesp.2015.10.002}.

\bibitem[Cumming(2014)]{cumming_new_2014}
Geoff Cumming.
\newblock The new statistics: Why and how.
\newblock \emph{Psychological science}, 25\penalty0 (1):\penalty0 7--29, 2014.
\newblock \doi{10.1177/0956797613504966}.

\bibitem[DeCoster et~al.(2011)DeCoster, Gallucci, and Iselin]{decoster_best_2011}
Jamie DeCoster, Marcello Gallucci, and Anne-Marie~R. Iselin.
\newblock Best practices for using median splits, artificial categorization, and their continuous alternatives.
\newblock \emph{Journal of Experimental Psychopathology}, 2\penalty0 (2):\penalty0 197--209, 2011.
\newblock \doi{10.5127/jep.008310}.

\bibitem[DeMaris(2002)]{demaris_explained_2002}
Alfred DeMaris.
\newblock Explained variance in logistic regression: A monte carlo study of proposed measures.
\newblock \emph{Sociological Methods \& Research}, 31\penalty0 (1):\penalty0 27--74, 2002.
\newblock \doi{10.1177/0049124102031001002}.

\bibitem[Dickey(1971)]{savage_dickey_1971}
James~M. Dickey.
\newblock {The Weighted Likelihood Ratio, Linear Hypotheses on Normal Location Parameters}.
\newblock \emph{The Annals of Mathematical Statistics}, 42\penalty0 (1):\penalty0 204 -- 223, 1971.
\newblock \doi{10.1214/aoms/1177693507}.

\bibitem[Gelman et~al.(2004)Gelman, Carlin, Stern, Vehtari, and Rubin]{bda2013}
Andrew Gelman, John~B. Carlin, Hal~S. Stern, Aki Vehtari, and Donald~B. Rubin.
\newblock \emph{Bayesian Data Analysis}.
\newblock {Chapman and Hall/CRC}, New York, NY, 3rd edition, 2004.
\newblock \doi{10.1201/b16018}.

\bibitem[Goodman et~al.(2016)Goodman, Fanelli, and Ioannidis]{goodman_reproducibility_2016}
Steven~N. Goodman, Daniele Fanelli, and John P.~A. Ioannidis.
\newblock What does research reproducibility mean?
\newblock \emph{Science Translational Medicine}, 8\penalty0 (341):\penalty0 341ps12--341ps12, 2016.
\newblock \doi{10.1126/scitranslmed.aaf5027}.

\bibitem[Gu et~al.(2018)Gu, Mulder, and Hoijtink]{gu_approximated_2018}
Xin Gu, Joris Mulder, and Herbert Hoijtink.
\newblock Approximated adjusted fractional bayes factors: A general method for testing informative hypotheses.
\newblock \emph{British Journal of Mathematical and Statistical Psychology}, 71\penalty0 (2):\penalty0 229--261, 2018.
\newblock \doi{10.1111/bmsp.12110}.

\bibitem[Hagle and Mitchell(1992)]{hagle_mitchell_goodness_1992}
Timothy~M. Hagle and Glenn~E. Mitchell.
\newblock Goodness-of-fit measures for probit and logit.
\newblock \emph{American Journal of Political Science}, 36\penalty0 (3):\penalty0 762--784, 1992.
\newblock \doi{10.2307/2111590}.

\bibitem[Hoijtink(2012)]{hoijtink_informative_2012}
Herbert Hoijtink.
\newblock \emph{Informative {H}ypotheses: {T}heory and {P}ractice for {B}ehavioral and {S}ocial {S}cientists}.
\newblock CRC Press, New York, NY, 2012.

\bibitem[Hoijtink(2021)]{hoijtink_prior_2021}
Herbert Hoijtink.
\newblock Prior sensitivity of null hypothesis bayesian testing.
\newblock \emph{Psychological Methods}, 2021.
\newblock \doi{10.1037/met0000292}.

\bibitem[Hoijtink et~al.(2019)Hoijtink, Mulder, van Lissa, and Gu]{hoijtink2019tutorial}
Herbert Hoijtink, Joris Mulder, Caspar van Lissa, and Xin Gu.
\newblock A tutorial on testing hypotheses using the {B}ayes factor.
\newblock \emph{Psychological methods}, 24\penalty0 (5):\penalty0 539, 2019.
\newblock \doi{10.1037/met0000201}.

\bibitem[Jeffreys(1961)]{jeffreys_1961}
Harold Jeffreys.
\newblock \emph{{Theory of probability}}.
\newblock Oxford University Press, Oxford, 3rd ed. edition, 1961.

\bibitem[Kass and Raftery(1995)]{kass_raftery_bayes_factors_1995}
Robert~E. Kass and Adrian~E. Raftery.
\newblock Bayes factors.
\newblock \emph{Journal of the American Statistical Association}, 90\penalty0 (430):\penalty0 773--795, 1995.
\newblock \doi{10.1080/01621459.1995.10476572}.

\bibitem[Kevenaar et~al.(2021)Kevenaar, Zondervan-Zwijnenburg, Blok, Schmengler, Fakkel, {de Zeeuw}, {van Bergen}, Onland-Moret, Peeters, Hillegers, Boomsma, and Oldehinkel]{kevenaar_bes_2021}
Sofieke~T. Kevenaar, Maria~A.J. Zondervan-Zwijnenburg, Elisabet Blok, Heiko Schmengler, M.~(Ties) Fakkel, Eveline~L. {de Zeeuw}, Elsje {van Bergen}, N.~Charlotte Onland-Moret, Margot Peeters, Manon~H.J. Hillegers, Dorret~I. Boomsma, and Albertine~J. Oldehinkel.
\newblock Bayesian evidence synthesis in case of multi-cohort datasets: An illustration by multi-informant differences in self-control.
\newblock \emph{Developmental Cognitive Neuroscience}, 47:\penalty0 100904, 2021.
\newblock \doi{10.1016/j.dcn.2020.100904}.

\bibitem[Klein et~al.(2014)Klein, Ratliff, Vianello, Adams, Bahník, Bernstein, Bocian, Brandt, Brooks, Brumbaugh, Cemalcilar, Chandler, Cheong, Davis, Devos, Eisner, Frankowska, Furrow, Galliani, Hasselman, Hicks, Hovermale, Hunt, Huntsinger, IJzerman, John, Joy-Gaba, Barry~Kappes, Krueger, Kurtz, Levitan, Mallett, Morris, Nelson, Nier, Packard, Pilati, Rutchick, Schmidt, Skorinko, Smith, Steiner, Storbeck, Van~Swol, Thompson, van~‘t Veer, Ann~Vaughn, Vranka, Wichman, Woodzicka, and Nosek]{klein_etal_replicability_2014}
Richard~A. Klein, Kate~A. Ratliff, Michelangelo Vianello, Reginald~B. Adams, {\v{S}}t{\v{e}}pán Bahník, Michael~J. Bernstein, Konrad Bocian, Mark~J. Brandt, Beach Brooks, Claudia~Chloe Brumbaugh, Zeynep Cemalcilar, Jesse Chandler, Winnee Cheong, William~E. Davis, Thierry Devos, Matthew Eisner, Natalia Frankowska, David Furrow, Elisa~Maria Galliani, Fred Hasselman, Joshua~A. Hicks, James~F. Hovermale, S.~Jane Hunt, Jeffrey~R. Huntsinger, Hans IJzerman, Melissa-Sue John, Jennifer~A. Joy-Gaba, Heather Barry~Kappes, Lacy~E. Krueger, Jaime Kurtz, Carmel~A. Levitan, Robyn~K. Mallett, Wendy~L. Morris, Anthony~J. Nelson, Jason~A. Nier, Grant Packard, Ronaldo Pilati, Abraham~M. Rutchick, Kathleen Schmidt, Jeanine~L. Skorinko, Robert Smith, Troy~G. Steiner, Justin Storbeck, Lyn~M. Van~Swol, Donna Thompson, A.~E. van~‘t Veer, Leigh Ann~Vaughn, Marek Vranka, Aaron~L. Wichman, Julie~A. Woodzicka, and Brian~A. Nosek.
\newblock Investigating variation in replicability.
\newblock \emph{Social Psychology}, 45\penalty0 (3):\penalty0 142--152, 2014.
\newblock \doi{10.1027/1864-9335/a000178}.

\bibitem[Klugkist and Hoijtink(2007)]{klugkist_bf_2007}
Irene Klugkist and Herbert Hoijtink.
\newblock The bayes factor for inequality and about equality constrained models.
\newblock \emph{Computational Statistics \& Data Analysis}, 51\penalty0 (12):\penalty0 6367--6379, 2007.
\newblock \doi{10.1016/j.csda.2007.01.024}.

\bibitem[Klugkist and Volker(2023)]{klugkist_volker_2023}
Irene Klugkist and Thom~Benjamin Volker.
\newblock {B}ayesian {E}vidence {S}ynthesis for informative hypotheses: An introduction.
\newblock \emph{Psychological Methods}, Advance online publication, 2023.
\newblock \doi{10.1037/met0000602}.

\bibitem[Klugkist et~al.(2005)Klugkist, Laudy, and Hoijtink]{klugkist_inequality_2005}
Irene Klugkist, Olav Laudy, and Herbert Hoijtink.
\newblock Inequality constrained analysis of variance: a bayesian approach.
\newblock \emph{Psychological methods}, 10\penalty0 (4):\penalty0 477--493, 2005.
\newblock \doi{10.1037/1082-989X.10.4.477}.

\bibitem[Kuiper et~al.(2013)Kuiper, Buskens, Raub, and Hoijtink]{kuiper_combining_2013}
Rebecca~M. Kuiper, Vincent Buskens, Werner Raub, and Herbert Hoijtink.
\newblock Combining statistical evidence from several studies: A method using bayesian updating and an example from research on trust problems in social and economic exchange.
\newblock \emph{Sociological Methods \& Research}, 42\penalty0 (1):\penalty0 60--81, 2013.
\newblock \doi{10.1177/0049124112464867}.

\bibitem[Lakens(2014)]{lakens2014performing}
Dani{\"e}l Lakens.
\newblock Performing high-powered studies efficiently with sequential analyses.
\newblock \emph{European Journal of Social Psychology}, 44\penalty0 (7):\penalty0 701--710, 2014.
\newblock \doi{10.1002/ejsp.2023}.

\bibitem[Lawlor et~al.(2017)Lawlor, Tilling, and Davey~Smith]{lawlor_triangulation_2017}
Debbie~A. Lawlor, Kate Tilling, and George Davey~Smith.
\newblock Triangulation in aetiological epidemiology.
\newblock \emph{International Journal of Epidemiology}, 45:\penalty0 1866–1886, 2017.
\newblock \doi{10.1093/ije/dyw314}.

\bibitem[Lipsey and Wilson(2001)]{lipsey_wilson_2001}
Mark~W Lipsey and David~B Wilson.
\newblock \emph{Practical {M}eta-{A}nalysis.}
\newblock SAGE Publications, Thousand Oaks, CA, 2001.

\bibitem[Lipton(2003)]{lipton2003inference}
Peter Lipton.
\newblock \emph{Inference to the best explanation}.
\newblock Routledge, New York, NY, 2003.

\bibitem[Ly et~al.(2016)Ly, Verhagen, and Wagenmakers]{ly_bf_2016}
Alexander Ly, Josine Verhagen, and Eric-Jan Wagenmakers.
\newblock Harold jeffreys’s default bayes factor hypothesis tests: Explanation, extension, and application in psychology.
\newblock \emph{Journal of Mathematical Psychology}, 72:\penalty0 19--32, 2016.
\newblock \doi{10.1016/j.jmp.2015.06.004}.

\bibitem[Lykken(1991)]{lykken_wrong_1991}
David~T. Lykken.
\newblock What's wrong with psychology, anyway?
\newblock In D.~Chiccetti and W.~Grove, editors, \emph{Thinking Clearly About Psychology}, pages 3--39. University of Minnesota Press, 1991.

\bibitem[Lynch(2007)]{lynch_introduction_2007}
Scott~M. Lynch.
\newblock \emph{Introduction to {Applied} {Bayesian} {Statistics} and {Estimation} for {Social} {Scientists}}.
\newblock Springer, New York, 2007.
\newblock \doi{10.1007/978-0-387-71265-9}.

\bibitem[Lynch and Bartlett(2019)]{lynch_bayesian_2019}
Scott~M. Lynch and Bryce Bartlett.
\newblock Bayesian statistics in sociology: Past, present, and future.
\newblock \emph{Annual Review of Sociology}, 45\penalty0 (1):\penalty0 47--68, 2019.
\newblock \doi{10.1146/annurev-soc-073018-022457}.

\bibitem[Mathison(1988)]{mathison1988triangulate}
Sandra Mathison.
\newblock Why triangulate?
\newblock \emph{Educational researcher}, 17\penalty0 (2):\penalty0 13--17, 1988.
\newblock \doi{10.3102/0013189X017002013}.

\bibitem[McKelvey and Zavoina(1975)]{mckelvey_zavoina_1975}
Richard~D. McKelvey and William Zavoina.
\newblock A statistical model for the analysis of ordinal level dependent variables.
\newblock \emph{The Journal of Mathematical Sociology}, 4\penalty0 (1):\penalty0 103--120, 1975.
\newblock \doi{10.1080/0022250X.1975.9989847}.

\bibitem[Mulder and Olsson-Collentine(2019)]{mulder_olssoncollentine_2019}
J.~Mulder and A.~Olsson-Collentine.
\newblock Simple bayesian testing of scientific expectations in linear regression models.
\newblock \emph{Behavior Research Methods}, 51\penalty0 (3):\penalty0 1117--1130, 2019.
\newblock \doi{10.3758/s13428-018-01196-9}.

\bibitem[Mulder(2014)]{mulder_prior_2014}
Joris Mulder.
\newblock Prior adjusted default bayes factors for testing (in)equality constrained hypotheses.
\newblock \emph{Computational Statistics \& Data Analysis}, 71:\penalty0 448--463, 2014.
\newblock \doi{10.1016/j.csda.2013.07.017}.

\bibitem[Mulder and Gu(2021)]{mulder_gu_bayesian_2021}
Joris Mulder and Xin Gu.
\newblock Bayesian testing of scientific expectations under multivariate normal linear models.
\newblock \emph{Multivariate Behavioral Research}, pages 1--29, 2021.
\newblock \doi{10.1080/00273171.2021.1904809}.

\bibitem[Mulder et~al.(2010)Mulder, Hoijtink, and Klugkist]{mulder_equality_2010}
Joris Mulder, Herbert Hoijtink, and Irene Klugkist.
\newblock Equality and inequality constrained multivariate linear models: Objective model selection using constrained posterior priors.
\newblock \emph{Journal of Statistical Planning and Inference}, 140\penalty0 (4):\penalty0 887--906, 2010.
\newblock \doi{10.1016/j.jspi.2009.09.022}.

\bibitem[Mulder et~al.(2021)Mulder, {van Lissa}, Williams, Gu, Olsson-Collentine, Boeing-Messing, and Fox]{BFpack}
Joris Mulder, Caspar {van Lissa}, Donald~R. Williams, Xin Gu, Anton Olsson-Collentine, Florian Boeing-Messing, and Jean-Paul Fox.
\newblock \emph{BFpack: Flexible Bayes Factor Testing of Scientific Expectations}, 2021.
\newblock URL \url{https://CRAN.R-project.org/package=BFpack}.
\newblock R package version 0.3.2.

\bibitem[Munaf{\`o} et~al.(2017)Munaf{\`o}, Nosek, Bishop, Button, Chambers, Percie~du Sert, Simonsohn, Wagenmakers, Ware, and Ioannidis]{munafo_manifesto_2017}
Marcus~R. Munaf{\`o}, Brian~A. Nosek, Dorothy V.~M. Bishop, Katherine~S. Button, Christopher~D. Chambers, Nathalie Percie~du Sert, Uri Simonsohn, Eric-Jan Wagenmakers, Jennifer~J. Ware, and John P.~A. Ioannidis.
\newblock A manifesto for reproducible science.
\newblock \emph{Nature Human Behaviour}, 1\penalty0 (1):\penalty0 0021, 2017.
\newblock \doi{10.1038/s41562-016-0021}.

\bibitem[Munafò and Smith(2018)]{munafo_robust_2018}
Marcus~R. Munafò and George~Davey Smith.
\newblock Robust research needs many lines of evidence.
\newblock \emph{Nature}, 553\penalty0 (7689):\penalty0 399--401, 2018.
\newblock \doi{10.1038/d41586-018-01023-3}.

\bibitem[Nosek et~al.(2012)Nosek, Spies, and Motyl]{nosek_scientific_2012}
Brian~A. Nosek, Jeffrey~R. Spies, and Matt Motyl.
\newblock Scientific utopia: Ii. restructuring incentives and practices to promote truth over publishability.
\newblock \emph{Perspectives on Psychological Science}, 7\penalty0 (6):\penalty0 615--631, 2012.
\newblock \doi{10.1177/1745691612459058}.

\bibitem[Nosek et~al.(2021)Nosek, Hardwicke, Moshontz, Allard, Corker, Dreber, Fidler, Hilgard, Struhl, Nuijten, Rohrer, Romero, Scheel, Scherer, Schönbrodt, and Vazire]{nosek_replicability_review_2021}
Brian~A. Nosek, Tom~E. Hardwicke, Hannah Moshontz, Aurélien Allard, Katherine~S. Corker, Anna Dreber, Fiona Fidler, Joe Hilgard, Melissa~Kline Struhl, Michèle~B. Nuijten, Julia~M. Rohrer, Felipe Romero, Anne~M. Scheel, Laura~D. Scherer, Felix~D. Schönbrodt, and Simine Vazire.
\newblock Replicability, robustness, and reproducibility in psychological science.
\newblock \emph{Annual Review of Psychology}, 73\penalty0 (1):\penalty0 null, 2021.
\newblock \doi{10.1146/annurev-psych-020821-114157}.

\bibitem[NWO(2020)]{nwo_replication_2020}
NWO.
\newblock Replication studies, Mar 2020.
\newblock URL \url{https://www.nwo.nl/en/researchprogrammes/replication-studies}.

\bibitem[O'Hagan(1995)]{ohagan_fractional_1995}
Anthony O'Hagan.
\newblock Fractional bayes factors for model comparison.
\newblock \emph{Journal of the Royal Statistical Society: Series B (Methodological)}, 57\penalty0 (1):\penalty0 99--118, 1995.
\newblock \doi{10.1111/j.2517-6161.1995.tb02017.x}.

\bibitem[{Open Science Collaboration}(2015)]{open_science_collab_2015}
{Open Science Collaboration}.
\newblock Estimating the reproducibility of psychological science.
\newblock \emph{Science}, 349\penalty0 (6251), 2015.
\newblock \doi{10.1126/science.aac4716}.

\bibitem[{R Core Team}(2021)]{R}
{R Core Team}.
\newblock \emph{R: A Language and Environment for Statistical Computing}.
\newblock R Foundation for Statistical Computing, Vienna, Austria, 2021.
\newblock URL \url{https://www.R-project.org/}.

\bibitem[Royall(1997)]{royall1997statistical}
Richard Royall.
\newblock \emph{{S}tatistical {E}vidence: A {L}ikelihood {P}aradigm}.
\newblock Routledge, New York, NY, 1997.

\bibitem[Schmidt(2009)]{schmidt_replication_2009}
Stefan Schmidt.
\newblock Shall we really do it again? the powerful concept of replication is neglected in the social sciences.
\newblock \emph{Review of General Psychology}, 13\penalty0 (2):\penalty0 90--100, 2009.
\newblock \doi{10.1037/a0015108}.

\bibitem[Sch{\"o}nbrodt et~al.(2017)Sch{\"o}nbrodt, Wagenmakers, Zehetleitner, and Perugini]{schonbrodt_sequential_2017}
Felix~D Sch{\"o}nbrodt, Eric-Jan Wagenmakers, Michael Zehetleitner, and Marco Perugini.
\newblock Sequential hypothesis testing with bayes factors: Efficiently testing mean differences.
\newblock \emph{Psychological methods}, 22\penalty0 (2):\penalty0 322, 2017.
\newblock \doi{10.1037/met0000061}.

\bibitem[Sutton and Abrams(2001)]{sutton_bayesian_meta2001}
Alex~J Sutton and Keith~R Abrams.
\newblock Bayesian methods in meta-analysis and evidence synthesis.
\newblock \emph{Statistical methods in medical research}, 10\penalty0 (4):\penalty0 277--303, 2001.
\newblock \doi{10.1177/096228020101000404}.

\bibitem[Tendeiro and Kiers(2019)]{tendeiro_kiers_2019}
Jorge~N. Tendeiro and Henk A.~L. Kiers.
\newblock A review of issues about null hypothesis bayesian testing.
\newblock \emph{Psychological Methods}, 24\penalty0 (6):\penalty0 774--795, 2019.
\newblock \doi{10.1037/met0000221}.

\bibitem[Trafimow et~al.(2018)Trafimow, Amrhein, Areshenkoff, Barrera-Causil, Beh, Bilgiç, Bono, Bradley, Briggs, Cepeda-Freyre, Chaigneau, Ciocca, Correa, Cousineau, de~Boer, Dhar, Dolgov, Gómez-Benito, Grendar, Grice, Guerrero-Gimenez, Gutiérrez, Huedo-Medina, Jaffe, Janyan, Karimnezhad, Korner-Nievergelt, Kosugi, Lachmair, Ledesma, Limongi, Liuzza, Lombardo, Marks, Meinlschmidt, Nalborczyk, Nguyen, Ospina, Perezgonzalez, Pfister, Rahona, Rodríguez-Medina, Romão, Ruiz-Fernández, Suarez, Tegethoff, Tejo, Van~de Schoot, Vankov, Velasco-Forero, Wang, Yamada, Zoppino, and Marmolejo-Ramos]{trafimow_manipulating_2018}
David Trafimow, Valentin Amrhein, Corson~N. Areshenkoff, Carlos~J. Barrera-Causil, Eric~J. Beh, Yusuf~K. Bilgiç, Roser Bono, Michael~T. Bradley, William~M. Briggs, Héctor~A. Cepeda-Freyre, Sergio~E. Chaigneau, Daniel~R. Ciocca, Juan~C. Correa, Denis Cousineau, Michiel~R. de~Boer, Subhra~S. Dhar, Igor Dolgov, Juana Gómez-Benito, Marian Grendar, James~W. Grice, Martin~E. Guerrero-Gimenez, Andrés Gutiérrez, Tania~B. Huedo-Medina, Klaus Jaffe, Armina Janyan, Ali Karimnezhad, Fränzi Korner-Nievergelt, Koji Kosugi, Martin Lachmair, Rubén~D. Ledesma, Roberto Limongi, Marco~T. Liuzza, Rosaria Lombardo, Michael~J. Marks, Gunther Meinlschmidt, Ladislas Nalborczyk, Hung~T. Nguyen, Raydonal Ospina, Jose~D. Perezgonzalez, Roland Pfister, Juan~J. Rahona, David~A. Rodríguez-Medina, Xavier Romão, Susana Ruiz-Fernández, Isabel Suarez, Marion Tegethoff, Mauricio Tejo, Rens Van~de Schoot, Ivan~I. Vankov, Santiago Velasco-Forero, Tonghui Wang, Yuki Yamada, Felipe C.~M. Zoppino, and Fernando Marmolejo-Ramos.
\newblock Manipulating the alpha level cannot cure significance testing.
\newblock \emph{Frontiers in Psychology}, 9, 2018.
\newblock \doi{10.3389/fpsyg.2018.00699}.

\bibitem[Van~de Schoot et~al.(2011)Van~de Schoot, Hoijtink, and Romeijn]{vandeschoot_informative_2011}
Rens Van~de Schoot, Herbert Hoijtink, and Jan-Willem Romeijn.
\newblock Moving beyond traditional null hypothesis testing: Evaluating expectations directly.
\newblock \emph{Frontiers in Psychology}, 2:\penalty0 24, 2011.
\newblock \doi{10.3389/fpsyg.2011.00024}.

\bibitem[Van~de Schoot et~al.(2017)Van~de Schoot, Winter, Ryan, Zondervan-Zwijnenburg, and Depaoli]{vandeschoot_systematic_2017}
Rens Van~de Schoot, Sonja~D Winter, Ois{\'\i}n Ryan, Mari{\"e}lle Zondervan-Zwijnenburg, and Sarah Depaoli.
\newblock A systematic review of {B}ayesian articles in psychology: The last 25 years.
\newblock \emph{Psychological Methods}, 22\penalty0 (2):\penalty0 217, 2017.
\newblock \doi{10.1037/met0000100}.

\bibitem[Volker(2022)]{volker_cooperation_2022}
Thom~Benjamin Volker.
\newblock The future is made today: Concerns for reputation foster trust and cooperation.
\newblock Master's thesis, Utrecht University, Department of Sociology, 2022.
\newblock Unpublished.

\bibitem[Wagenmakers et~al.(2018)Wagenmakers, Marsman, Jamil, Ly, Verhagen, Love, Selker, Gronau, {\v{S}}m{\'i}ra, Epskamp, Matzke, Rouder, and Morey]{Wagenmakers_bayesian_2018}
Eric-Jan Wagenmakers, Maarten Marsman, Tahira Jamil, Alexander Ly, Josine Verhagen, Jonathon Love, Ravi Selker, Quentin~F. Gronau, Martin {\v{S}}m{\'i}ra, Sacha Epskamp, Dora Matzke, Jeffrey~N. Rouder, and Richard~D. Morey.
\newblock Bayesian inference for psychology. part i: Theoretical advantages and practical ramifications.
\newblock \emph{Psychonomic Bulletin {\&} Review}, 25\penalty0 (1):\penalty0 35--57, 2018.
\newblock \doi{10.3758/s13423-017-1343-3}.

\bibitem[Zellner and Siow(1980)]{zellner_siow_1980}
A.~Zellner and A.~Siow.
\newblock Posterior odds ratios for selected regression hypotheses.
\newblock \emph{Trabajos de Estadistica Y de Investigacion Operativa}, 31\penalty0 (1):\penalty0 585--603, 1980.
\newblock \doi{10.1007/BF02888369}.

\bibitem[Zondervan-Zwijnenburg et~al.(2020{\natexlab{a}})Zondervan-Zwijnenburg, Veldkamp, Neumann, Barzeva, Nelemans, van Beijsterveldt, Branje, Hillegers, Meeus, Tiemeier, Hoijtink, Oldehinkel, and Boomsma]{zondervan_parental_2019}
M.~A.~J. Zondervan-Zwijnenburg, Sabine~A.M. Veldkamp, Alexander Neumann, Stefania~A. Barzeva, Stefanie~A. Nelemans, Catharina~E.M. van Beijsterveldt, Susan~J.T. Branje, Manon~H.J. Hillegers, Wim~H.J. Meeus, Henning Tiemeier, Herbert~J.A. Hoijtink, Albertine~J. Oldehinkel, and Dorret~I. Boomsma.
\newblock Parental age and offspring childhood mental health: A multi-cohort, population-based investigation.
\newblock \emph{Child Development}, 91\penalty0 (3):\penalty0 964--982, 2020{\natexlab{a}}.
\newblock \doi{10.1111/cdev.13267}.

\bibitem[Zondervan-Zwijnenburg et~al.(2020{\natexlab{b}})Zondervan-Zwijnenburg, Richards, Kevenaar, Becht, Hoijtink, Oldehinkel, Branje, Meeus, and Boomsma]{zondervan_robust_2020}
M.A.J. Zondervan-Zwijnenburg, J.S. Richards, S.T. Kevenaar, A.I. Becht, H.J.A. Hoijtink, A.J. Oldehinkel, S.~Branje, W.~Meeus, and D.I. Boomsma.
\newblock Robust longitudinal multi-cohort results: The development of self-control during adolescence.
\newblock \emph{Developmental Cognitive Neuroscience}, 45:\penalty0 100817, 2020{\natexlab{b}}.
\newblock \doi{10.1016/j.dcn.2020.100817}.

\end{thebibliography}

\end{document}